\documentclass[11pt]{article}

\pdfoutput=1

\usepackage[top=80pt,bottom=85pt,left=70pt,right=70pt]{geometry}
\usepackage{amssymb}
\usepackage{amsmath}
\usepackage{graphicx,color,subfigure}
\usepackage{float}
\usepackage{cite}
\usepackage{enumerate}
\usepackage[debug,pageanchor=false]{hyperref}
\definecolor{link}{rgb}{.8,.15,.1}
\hypersetup{colorlinks=true,linkcolor=link,citecolor=link,urlcolor=link,linktocpage}
\usepackage{tikz}
\usetikzlibrary{arrows,decorations.pathmorphing,backgrounds,positioning,fit,petri}

\DeclareMathOperator{\g}{\gamma}
\DeclareMathOperator{\eps}{\epsilon}

\makeatletter
\@addtoreset{equation}{section}
\makeatother

\newcommand{\beq}{\begin{equation}}
\newcommand{\eeq}{\end{equation}}
\newcommand{\bea}{\begin{eqnarray}}
\newcommand{\eea}{\end{eqnarray}}
\newcommand{\nn}{\nonumber}

\begin{document}

\begin{titlepage}

\begin{center}

\vskip .5in %.3in 
\noindent

{\Large \bf{Mink$_4\times S^2$ Solutions of 10 and 11 Dimensional Supergravity}}

\bigskip\medskip

Andrea Legramandi$^{a}$ and Niall T. Macpherson$^{b}$\\

\bigskip\medskip
{\small 

$a$: Dipartimento di Fisica, Universit\`a di Milano--Bicocca, \\ Piazza della Scienza 3, I-20126 Milano, Italy \\ and \\ INFN, sezione di Milano--Bicocca\\[1mm]

$b$:	 SISSA International School for Advanced Studies\\
Via Bonomea 265, 34136 Trieste \\
			and\\ INFN, sezione di Trieste
	
}

\vskip .5cm %.3cm
{\small \tt nmacpher@sissa.it, a.legramandi@campus.unimib.it}
\vskip .9cm %.6cm
     	{\bf Abstract }

\vskip .1in
\end{center}

\noindent
We complete the classification of Mink$_4$ solutions preserving $\mathcal{N}=2$ supersymmetry and SU(2) R-symmetry parameterised by a round $S^2$ factor. We  consider eleven-dimensional supergravity  and relax the assumptions of earlier works in type II theories. We show that, using chains of dualities, all solutions of this type can be generated from one of two master classes: an SU(2)-structure in M-theory and a conformal Calabi--Yau in type IIB. Finally, using our results, we recover AdS$_5\times S^2$ solutions in M-theory and construct a compact Minkowski solution with Atiyah--Hitchin singularity.

\noindent
 
\vfill
\eject

\end{titlepage}

\tableofcontents

\section{Introduction}
Some of the most physically relevant solutions in supergravity are those exhibiting a warped Minkowski factor. From early on, the main reason for this is that Minkowski vacua of string and M-theory are required to make contact with known particle physics phenomena in  four dimensions, where one should arrange for the co-dimensions to be compact. Another reason, which clearly gained considerable traction with the advent of the AdS/CFT correspondence, is that all AdS solutions admit a description in terms of a foliation of Minkowski over a non compact interval - namely the Poincar\'e patch.

The most simple way to realise a Minkowski vacua from ten or eleven dimensions is to assume that the compact internal space accommodates some holonomy group, specifically SU(3) or G$_2$ for compactifications of string theory or M-theory down to four dimensions respectively. Such solutions preserve (at least) $\mathcal{N}=2$ supersymmetry and have been well studied in the literature \cite{Candelas:1985en,Witten:1997sc,Acharya:2001gy,Acharya:2004qe,Douglas:2015aga}, with a resurgence of interest in the  G$_2$ case  in recent years 
\cite{Acharya:2015oea,Kennon:2018eqg,Andriolo:2018yrz,Fiset:2018huv} (see also \cite{delaOssa:2014lma,Fiset:2017auc,delaOssa:2017pqy,delaOssa:2017gjq}
 for G$_2$ arising in a heterotic context).  However, such manifolds support neither fluxes nor a warping of the Minkowski directions, so it is reasonable to expect that they represent a rather small region of the space of possible solutions.  The inclusion of fluxes requires one to generalise the notion of holonomy group to structure group (or G-structure) 	\cite{Gauntlett:2001ur,Friedrich:2001yp,Gauntlett:2002sc,Gauntlett:2003cy}. 
 %The G-structure the internal space will support is SU(3)$\times$ SU(3) (usually characterised by the largest subgroup common to both factors in the product) in type II supergravity or an SU(3)-structure in eleven dimensions. 
 It is well known that no compact regular solutions of this type exist \cite{Gibbons:1984kp,deWit:1986mwo,Maldacena:2000mw} - indeed a necessary element of such constructions are localised singularities, namely O-planes and their generalisations through string dualities, or lifts to M-theory (e.g. Atiyah--Hitchin singularities). Most progress constructing compact solutions with fluxes has happened within another restrictive ansatz, with the internal space assumed to be conformally a holonomy manifold \cite{Becker:1996gj,Dasgupta:1999ss,Giddings:2001yu}, allowing one to broadly use the same mathematical tools as before. At this point all  Mink$_d$ solutions for $d>1$ (with the exception of Mink$_2$ in eleven dimensions) have been classified (see \cite{Martelli:2003ki,Kaste:2003zd,DallAgata:2003txk,Lukas:2004ip,Gauntlett:2004zh} for eleven dimensions, \cite{Grana:2004bg,Grana:2005sn,Haack:2009jg,Lust:2010by,Prins:2013koa,Prins:2013wza,Rosa:2013lwa,Macpherson:2017mvu} for ten dimensions), however finding solutions beyond the ansatz of warped holonomy has proved challenging - see 	\cite{Candelas:2014jma, Candelas:2014kma} for success in this direction. The issue appears to be one of tractability, so it would be helpful to have some additional guiding principle.

AdS solutions necessitate the inclusion of fluxes, early examples were also constructed by studying warped holonomy manifolds (albeit now non compact) using the Freund-Rubin ansatz \cite{Freund:1980xh}. By now the state of affairs for AdS solutions beyond this class is rather more developed than for Minkowski. Many AdS solutions preserving 16 supercharges or more, when they exist, are either completely known (see the cases of  AdS$_7$ \cite{Apruzzi:2013yva}, AdS$_6$ \cite{Passias:2012vp} in IIA/M-theory AdS$_5\times S^2$ \cite{Colgain:2011hb} in IIB\footnote{There is only the $\mathbb{Z}_k$ orbifold of AdS$_5\times S^5$.}), or known locally up to solving comparatively simple partial differential equations (see the  cases of AdS$_6$ \cite{Apruzzi:2014qva,DHoker:2016ujz} and AdS$_4$ \cite{DHoker:2016ujz} in IIB and AdS$_5\times S^2$ in M-theory/IIA\footnote{see \cite{ReidEdwards:2010qs} for some explicit examples in IIA.} \cite{Lin:2004nb,Macpherson:2016xwk}). The exceptions are AdS$_{2/3}$ (see \cite{Corbino:2017tfl,Dibitetto:2018gbk,Dibitetto:2018ftj} for certain ans$\ddot{a}$tze\footnote{Recently there was also \cite{Legramandi:2018qkr} which makes no assumption beyond the existence of a time-like Killing vector.}), AdS$_4$ in IIA/M-theory, and AdS$_5$ with the requisite R-symmetry not realised by a round $S^2$ - but these will not be our focus here. Many cases with less supersymmetry are also well studied, in particular, with the recent addition of \cite{Dibitetto:2018ftj}, all AdS$_d$ solutions with $d>2$ and at least minimal supersymmetry have been classified (see \cite{Martelli:2003ki,Grana:2005sn,Gauntlett:2004zh,Gauntlett:2005ww,Lust:2004ig,Koerber:2008rx,Passias:2018zlm,Dibitetto:2018ftj,Apruzzi:2015zna,Couzens:2016iot}, and the AdS$_{6,7}$ classifications above\footnote{We don't speak of $d>7$ because such AdS solutions break SUSY, however see  \cite{Cordova:2018eba} for examples of AdS$_8$.}) - broadly speaking these classes have been more successfully used to find solutions beyond restrictive ans$\ddot{a}$tze than their Minkowski counter-parts. The AdS/CFT obviously motivated these classifications, but in addition to the extra impetuous this provided, these classifications benefit from additional symmetry with respect to the Minkowski cases. One issue however, is most classifications for AdS, while quite detailed, assume global AdS factors from the start, so are not particularly useful for studying certain non conformal behaviors such as RG flows. For this reason it would be useful to embed the AdS classes into a more general set up\footnote{See \cite{Malek:2018zcz} and \cite{DeLuca:2018zbi} for some  recent progress in this direction  using exceptional field theory and consistent truncation respectively.}.

Here and in the earlier works of \cite{Macpherson:2016xwk}, \cite{Apruzzi:2018cvq}, the philosophy is to learn from the successes of the AdS classifications and perform a Mink$_4$ classification that assumes some additional symmetry so as to enable a more detailed description than previous efforts. A good starting point for this is to classify  $\mathcal{N}=2$ that preserve an SU(2) R-symmetry in the form of a round $S^2$ factor (see also \cite{Macpherson:2017mvu} where Mink$_3\times S^3$ are classified) - one can then try to break some of this (super)symmetry and generate many more solutions in the spirit of \cite{Rota:2015aoa}. An SU(2) R-symmetry is a necessary part of the super-conformal algebra in $d=5,6$ and $d=4$ with $\mathcal{N}=2$. As such a corollary to this classification endeavor is that it provides an embedding of the known half-BPS AdS$_d$ solutions (with $d>4$) into a  broader context still preserving SU$(2)_R$.   In \cite{Macpherson:2016xwk}, \cite{Apruzzi:2018cvq} such  classifications where performed in types IIA and IIB respectively, under a certain simplifying assumptions. The purpose of this work is to complete this program by classifying solutions in eleven-dimensional supergravity and relaxing the assumptions made in the earlier works.\\

The lay out of the paper is as follows: In section \ref{sec:Mtheoryclassification} we classify $\mathcal{N}=2$ Mink$_4$ solutions in M-theory that realise an SU(2) R-symmetry in terms of a round $S^2$  factor in there internal space, which leaves a five-manifold to be determined by geometric supersymmetry constraints. We find that there are two classes of solution, we refer to as A and B. Class A is governed by an SU(2)-structure in five dimensions while Class B is the M5-branes with SO(3) rotational invariance in it's co-dimensions. The physical fields of case B also solve the supersymmetry conditions of case A, however the Killing spinor and G-structure vielbein are different, and this difference can become physical upon reduction to IIA. Additionally we make some simple ans$\ddot{a}$tze  for Case A, specifically a conformal CY2 ansatz and one where the internal space contains an additional squashed $S^3$.

In section \ref{sec:nonequalnorm} we turn our attention to Mink$_4\times S^2$ in ten dimensions. We drop the assumption of equal Majorana--Weyl spinor norm made in \cite{Macpherson:2016xwk}, \cite{Apruzzi:2018cvq} making the classification of type II completely general. We find that all solutions that lie outside the existing classifications always have a $\text{U}(1)\times \text{U}(1)$ flavour symmetry and can be generated from a ``parent'' system in M-theory (the case B M5-brane) via  chains of dualities. 

Section \ref{sec:master} elucidates the connection between the different classes of solution contained in \cite{Macpherson:2016xwk}, \cite{Apruzzi:2018cvq}. We are able to show that all such  Mink$_4\times S^2$ solutions in type II can be generated from two master systems using chains of dualities - a conformal Calabi--Yau system in IIB and the SU(2)-structure in in M-theory (Case A). 

Finally in section \ref{sec:ex} we close with some simple examples contained within M-theory case A. Specifically we establish how the $\mathcal{N}=2$ AdS$_5$ class in M-theory is embedded in the SU(2)-structure of case A and we point the way towards some compact Mink$_4$ solutions with fluxes and Atiyah--Hitchin singularities.

\section{Mink$_4\times S^2$ in M-theory}\label{sec:Mtheoryclassification}
In this section we will classify $\mathcal{N}=2$ supersymmetric warped Mink$_4$ solutions in eleven dimensions realising an SU(2) R-symmetry with a round $S^2$. 
\subsection{The spinor ansatz}
We are interested in supersymmetric solutions to eleven-dimensional supergravity with a warped four-dimensional Minkowski factor. As such we decompose the metric as
\beq
\label{fluxdef}
ds^2= e^{2\Delta}ds^2(\mathbb{R}_{1,3})+ ds^2(M_7),
\eeq
where $e^{2\Delta}$ is a function with support on $M_7$ only and the four-form flux $G$ is necessarily purely magnetic.  As we seek $\mathcal{N}=2$ solutions respecting the warped product  $\mathbb{R}_{1,3}\times M_7$, our eleven-dimensional spinor will decompose as
\beq\label{eq:Neqspinors}
\epsilon= \sum_{a=1}^2 \bigg(\zeta^a_+\otimes \chi^a+ (\zeta^a_+)^c\otimes (\chi^a)^c\bigg)
\eeq
where $ \zeta^a_+$ is a doublet of positive chirality Mink$_4$ spinors, $\chi^a$ a doublet of seven-dimensional spinors such that
\begin{equation}
\label{7dspinorconditions}
|| \chi^a ||^2 = e^{\Delta}, \qquad \overline{\chi^a} \chi^a = 0,
\end{equation}
 and $c$ denotes Majorana conjugation. It is argued in \cite{Lukas:2004ip}, that it is possible to be slightly more general than this, however we show in Appendix \ref{sec: noLukas} that \eqref{eq:Neqspinors} is sufficient for our considerations. We will also  assume that our spinors $\chi^a$ are charged under an SU$(2)$ R-symmetry, realised by a round $S^2$ factor in the internal geometry so that the metric and flux on $M_7$ further decompose as
\beq
\label{eq:7dfluxesansatz}
ds^2(M_7)= e^{2C}ds^2(S^2)+ ds^2(M_5),~~~~ G= F_4+ e^{2C}\text{Vol}(S^2)\wedge F_2,
\eeq
i.e., as a foliation of $S^2$ over $M_5$, where $e^{2C}$ is a function depending on the coordinates on the five-dimensional manifold.
At the level of the spinors the SU$(2)$ R-symmetry will be realised by decomposing our doublet of spinors as
\beq
\chi^a= \xi^a \otimes \eta^1+ \hat\xi^a \otimes \eta^2,
\eeq
where $\xi^a$ and $\hat\xi^a$ are two SU$(2)$ doublets formed from the Killing spinors on $S^2$, $\xi$, which obeys 
\beq\label{eq:S2KSE}
\nabla_{\mu}\xi=\frac{i}{2}\sigma_\mu\xi,
\eeq
where we used the Pauli matrices to represent the Clifford algebra on $S^2$.
As established in \cite{Macpherson:2016xwk}, the doublets take the form
\beq
\xi^a= \left(\begin{array}{c}\xi\\\xi^c \end{array}\right),\quad \hat\xi^a= \left(\begin{array}{c}\sigma_3\xi\\ \sigma_3\xi^c \end{array}\right)
\eeq
where $\sigma_3$ is the 2d chirality matrix and $\xi^c= \sigma_2 \xi^{*}$ is the Majorana conjugate. Plugging $\epsilon$ into the the eleven-dimensional Killing spinor equation we get two independent systems of equations for the seven dimensional spinors $\chi^1$ and $\chi^2$ respectively. However, given the doublet transformation property under the spinorial Lie derivative
\beq
\mathcal{L}_{K_i}\xi^a= \frac{i}{2}(\sigma^i)^a_{~b}\xi^b,~~~~\mathcal{L}_{K_i}\hat\xi^a= \frac{i}{2}(\sigma^i)^a_{~b}\hat\xi^b,
\eeq
where $K_i$ are the Killing vectors of SU$(2)$, one finds that each system is implied by the other through this map, provided a solution respects the SU(2) R-symmetry. As such it is sufficient to solve for just one $\mathcal{N}=1$ sub-sector governed by for instance $\chi^1$, and impose that all physical fields of a solution depend on the invariant forms of SU(2), then the second $\mathcal{N}=1$ sub-sector governed by $\chi^2$ is implied.

In the following section we shall establish a set of geometric conditions for the five-dimensional submanifold $M_5$, that are nessisary and sufficient for supersymmetry.
\subsection{Supersymmetry conditions from seven to five dimensions}
The $\mathcal{N}=1$ supersymmetry conditions for a warped product of $\mathbb{R}_{1,3}\times M_7$ with spinors of the form
\beq
\epsilon= \zeta_+ \otimes \chi+ \text{m.c.}
\eeq
were studied in \cite{Kaste:2003zd,DallAgata:2003txk}, where necessary and sufficient geometric conditions were derived for the preservation of supersymmetry in the particular case $\overline{\chi}\chi = 0$. In the convention of \cite{Gauntlett:2004zh} these are
\begin{subequations}
\label{eqconditions}
\begin{align}
&d(e^{2\Delta}   K) = 0 ,  \label{Veq}\\
&d(e^{4\Delta} J) = -e^{4\Delta} \star_7 G,  \label{Jeq}\\
&d(e^{3\Delta} \Omega ) = 0 ,   \label{Omegaeq}\\
&d(e^{2\Delta} J \wedge J) = - 2 e^{2 \Delta}  G \wedge K\label{JJeq}
\end{align}
\end{subequations}
where 
\beq\label{eq:SU3eqs}
e^{\Delta}=\chi^{\dag}\chi,~~~e^{\Delta}K_a=\chi^{\dag}\gamma_a\chi,\quad e^{\Delta}J_{ab}=-i\chi^{\dag}\gamma_{ab}\chi,\quad e^{\Delta}\Omega_{abc}=-i\chi^{c\dag}\gamma_{abc}\chi,
\eeq
define an SU$(3)$-structure on the 7 dimensional internal space.
In fact these geometric conditions combined with Bianchi identity of $G_4$ are necessary and sufficient conditions for a solution with $\mathcal{N}=1$ supersymmetry to exist \cite{Gauntlett:2002fz}.

Since we seek a solution respecting the SU$(2)$ isometry of $S^2$, we can extract conditions on $M_5$ from \eqref{Veq}-\eqref{JJeq} by imposing \eqref{eq:7dfluxesansatz} and that our seven-dimensional spinor  takes the form
\beq\label{eq:7dspinor}
\chi= e^{\frac{\Delta}{2}}(\xi\otimes \eta^1+ \sigma_3\xi\otimes \eta^2),
\eeq
where $\eta^i$ and $\Delta$ are independent of the $S^2$ directions.

The first condition we encounter comes from the left hand expression in \eqref{7dspinorconditions}, which given \eqref{eq:7dspinor} yields
\begin{equation}
\chi^{\dag}\chi=e^{\Delta}(\eta_1^\dagger \eta_1 + \eta_2^\dagger \eta_2  ) + y_3 e^{\Delta}(\eta_1^\dagger \eta_2 + \eta_2^\dagger \eta_1  ) .
\end{equation}
This is only  consistent with $e^{\Delta}=\chi^{\dag}\chi$ being an SU(2) singlet if
\begin{equation}
\label{eq0form}
|| \eta_1||^2 + ||\eta_2||^2  =  1 , \qquad \text{Re}(\eta_1^\dagger \eta_2 ) = 0.
\end{equation}
For similar reasons the right hand expression of \eqref{7dspinorconditions} imposes
\begin{equation}
\label{eq0form2}
\overline{\eta_1} \eta_2 = 0 .
\end{equation}
Let's see how the seven dimensional bi-linears decompose into products of two- and five-dimensional ones. To do this we use the following gamma-matrix representation
\beq
\label{eq:gamma7d_rep}
\gamma^{(7)}_i= e^{C}\sigma_a\otimes\mathbb{I}_4,~~~~\gamma^{(7)}_a= \sigma_3\otimes \sigma_a,~~~ B_7= \sigma_2\otimes \sigma_1\otimes \sigma_2.
\eeq 
We also decompose the five-dimensional spinors in a common basis in terms of a unit norm spinor $\eta$ as
\beq
\eta_1= q_1 \eta,\quad \eta_2 = q_2(i\cos\alpha  \eta+ \frac{1}{2}\sin\alpha\overline{w} \eta),~~~~q_1^2+ q_2^2 =1,
\eeq
which is the most general parametrisation consistent with \eqref{eq0form}, \eqref{eq0form2}.
To calculate the forms in \eqref{eq:SU3eqs} we will make repeated use of the bilinear product identity
\beq
\big[\xi^1\otimes \eta^1\big]\otimes \big[\xi^2\otimes \eta^{2}\big]^\dag = (\eta^1\otimes \eta^{2\dag})_+\wedge (\xi^1\otimes\xi^{2\dag}) +(\eta^1\otimes \eta^{2\dag})_-\wedge (\sigma_3\xi^1\otimes\xi^{2\dag})
\eeq
where $\pm$ denotes the even/odd degree components of a form only, while the presence of $\sigma_3$ depends on our parametrisation of the gamma matrices \eqref{eq:gamma7d_rep}. The bi-linears that follow from $\eta$ are given in \cite{Apruzzi:2015zna} and read: 
\begin{align}
\eta\otimes \eta^{\dag}&= \frac{1}{4}(1+v)\wedge e^{-i j_2},\quad \eta\otimes \eta^{c\dag}= \frac{1}{4}(1+v)\wedge \omega_2,\nn\\[2mm]
\omega_2&= w\wedge u,~~~~ j_2=\frac{i}{2}(w\wedge \overline{w}+u\wedge \overline{u}),
\end{align}
where 
\beq
v,~w_1=\text{Re}w,~w_2=\text{Im}w~u_1=\text{Re}u,~u_2=\text{Im}u
\eeq
defines a vielbein in five dimensions. And finally the bi-linears that follow from $\xi$ are \cite{Macpherson:2016xwk}, 
\begin{align}
\xi\otimes\xi^{\dag}&=\frac{1}{2}(1+ k_3-i y_3 \text{Vol}(S^2)),\quad \xi\otimes\xi^{c\dag}=-\frac{1}{2}(k_1+ i k_2- i (y_1+ i y_2)\text{Vol}(S^2)),\nn\\[2mm]
\sigma_3\xi\otimes\xi^{\dag}&=\frac{1}{2}(y_3+ i dy_3- i \text{Vol}(S^2)),\quad \sigma_3\xi\otimes\xi^{c\dag}=-\frac{1}{2}( y_1+ i y_2-i d(y_1+ i y_2),
\end{align}
where $y_i$ are coordinates embedding $S^2$ into $\mathbb{R}^3$ and $k_i$ are one forms dual to the Killing vectors of SU$(2)$ which may be parameterised as
\beq
k_i= \epsilon_{ijk}y_j dy_k.
\eeq
The first thing we calculate is the one form
\begin{align}\label{eq:opriginalK}
K &=- q_1q_2e^{C}\cos\alpha dy_3+ q_1q_2\sin\alpha w_1+ q_2^2y_3\cos\alpha(\cos\alpha v- \sin\alpha w_2)\nn\\[2mm]
&+ \frac{1}{2}(q_1^2-q_2^2)( k_3+ y_3 u_2).
\end{align}
The appearance of $k_3$ here is opportune because $dk_i = 2y_i \text{Vol}(S^2)$ which means that the only way to make $K$ consistent with \eqref{Veq} is to set the coefficient of $k_3$ to zero - thus we can without loss of generality take
\beq
q_1=q_2= \frac{1}{\sqrt{2}},
\eeq
and then rotate the five-dimensional frame such that \eqref{eq:opriginalK} becomes simply
\beq\label{eq:K}
K =- e^{C}\cos\alpha \, dy_3+ \sin\alpha \, w_1+ y_3\cos\alpha \, v.
\eeq
In this frame the other forms become
\begin{align}
J&=e^{C}(\sin\alpha dy_3\wedge w_2- v\wedge k_3) + y_3(\sin\alpha w_2\wedge V+e^{2C}\text{Vol}(S^2))\nn\\[2mm]
&+\cos\alpha w_1\wedge w_2+ u_1\wedge u_2, \nn\\[2mm]
\Omega&= e^{C}u\wedge\big(i\sin\alpha  v\wedge (k_1+ i k_2)-(\cos\alpha w_1+ i w_2)\wedge (y_1+ i dy_2)\big) ,\nn\\[2mm]
\frac{1}{2}J\wedge J&= e^{2C}(\sin\alpha  w_2\wedge v+ y_3 (\cos\alpha w_!\wedge w_2+ u_1\wedge u_2)\wedge \text{Vol}(S^2)\nn\\[2mm]
& -e^{C}\big(w_2\wedge u_1\wedge u_2\wedge dy_3+(\cos\alpha w_1\wedge w_2+ u_1\wedge u_2)\wedge v\wedge k_3\big)\nn\\[2mm] 
&+ \cos\alpha w_1\wedge w_2\wedge u_1 \wedge u_2+ y_3 \sin\alpha w_2\wedge u_1\wedge u_2\wedge v\label{eq:M-theroy forms}
\end{align}
where in simplifying the last of these we make use of the identity
\beq
 k_{(i)}\wedge d y_{(i)}+ y_{(i)}^2 \text{Vol}(S^2) = \text{Vol}(S^2).
\eeq
We now plug \eqref{eq:M-theroy forms} into \eqref{Veq}-\eqref{JJeq} and factor out all dependence on $S^2$. The result of this operation is the following set of five-dimensional form constraints
\begin{subequations}
\begin{align}
&\sin\alpha d(e^{2\Delta} w)=\sin\alpha d(e^{- \Delta} u)=d(e^{2\Delta+C})+ e^{2\Delta} v=d\alpha=0,\label{eq: 5dsystem1}\\[2mm]
&d(e^{\Delta}(\cos\alpha w_1+ i w_2)\wedge u)= F_4 =0,\label{eq: 5dsystem2}\\[2mm]
&\sin\alpha(d(e^{2\Delta+2C}w_2\wedge v)+ e^{2\Delta+2C}w_1\wedge F_2)=0,\label{eq: 5dsystem3}\\[2mm]
&d(e^{-2\Delta}(\cos\alpha w_1\wedge w_2+ u_1\wedge u_2)+ e^{-2\Delta}\cos\alpha F_2\wedge v=0,\label{eq: 5dsystem4}\\[2mm]
&d(e^{4\Delta}(\cos\alpha w_1\wedge w_2+ u_1\wedge u_2)=e^{4\Delta}\star_5 F_2.\label{eq: 5dsystem5}
\end{align}
\end{subequations}
Clearly the behavior is quite different depending on whether or not $\alpha=0$, so one should look at these cases separately, which we now proceed to do in the next section. Note that solving these conditions and the Bianchi identity for the flux, implies the rest of the equations of motion \cite{Prins:2013wza}.

Before we move on let briefly examine the form of the SU(3)-structure in the internal six dimensions orthogonal to $K$. Any SU(3)-structure can be expressed in canonical form in terms  of a complex vielbein $E^i$ as
\begin{align}
J&=\frac{i}{2}\big(E^1\wedge \overline{E}^1+E^2\wedge \overline{E}^2+E^3\wedge \overline{E}^3\big),\nn\\[2mm]
\Omega&=E^1\wedge E^2\wedge E^3.
\end{align}
With a little effort, one can reverse engineer a complex vielbein  by manipulating \eqref{eq:M-theroy forms}, we find 
\begin{align}
E^1= \cos\alpha w_1+ i w_2 +\sin\alpha(e^{C} dy_3- y_3 v),~~~ E^2=u,~~~E^3=(y_1+i y_2) v- e^{C}d(y_1+ i y_2),\nn
\end{align}
where it is easy to confirm that this does indeed lead to a factorised metric $M_7= S^2\times M_5$
\beq
K^2+ E^i \overline{E}^i= e^{2C}\text{Vol}(S^2)+v^2+ w_1^2+ w_2^2+ u_1^2+ u_2^2.
\eeq
We shall now proceed to classify the solutions that follow from the five-dimensional supersymmetry conditions, \eqref{eq: 5dsystem1}-\eqref{eq: 5dsystem5} - there are two cases contained in sections \ref{sec: CaseA} and \ref{sec: CaseB}.

\subsection{Case A: $\alpha = 0$, SU(2)-structure}\label{sec: CaseA}
In this case $\eta_1$ and $\eta_2$ are proportional to one another, specifically  $\eta_1= i \eta_2$, and so they define a SU$(2)$-structure on $M_5$,
\beq\label{eq:SU2veil}
\omega_2 = u\wedge w,\quad j=\frac{i}{2}(u\wedge \overline{u} + w \wedge \overline{w}).
\eeq
This means that we cannot find local expression for the vielbein $u,w$ without any further assumptions.
The purely geometric supersymmetry conditions are simply
\begin{align}
&d(e^{2 \Delta + C}) + e^{2 \Delta} v = 0\label{SU2susy1},\\[2mm]
&d(e^\Delta \omega_2) = 0,\label{SU2susy2}
\end{align}
while those involving the flux read
\begin{equation}
\label{1casefluxcondition}
d(e^{4 \Delta} j_2 ) - e^{4 \Delta} \star_5 F_2 = d(e^{-2 \Delta} j_2) + e^{-2 \Delta} F_2 \wedge v = 0.
\end{equation}
We have a solution whenever we solve \eqref{SU2susy1}, \eqref{SU2susy1}, \eqref{1casefluxcondition} and the Bianchi identity of $G$ is satisfied, which requires
\beq
d(e^{2C} F_2) =  0,
\eeq
away from localised sources\footnote{The inclusion of such sources would put delta-function sources on the left hand side of the equality.}. We can solve \eqref{SU2susy2} without loss of generality by introducing a local coordinate $\rho$ such that
\beq
\rho = e^{2 \Delta + C},\quad  v=-e^{-2 \Delta}d\rho.
\eeq
Solutions in this class then take the form
\begin{align}\label{eq:ClassAsol}
ds^2&= e^{2\Delta}ds^2(\text{Mink}_4)+e^{-4\Delta}\big(d\rho^2+\rho^2ds^2(S^2)\big)+ ds^2(\text{M}_4),\nn\\[2mm]
G&=  e^{-8 \Delta}\rho^2\text{Vol}(S^2)\wedge\star_5 d(e^{4 \Delta} j_2 )
\end{align}
where $M_4$ supports the SU(2)-structure, that given \eqref{SU2susy2} is complex.

In the next subsections we shall make some assumptions about the form of the SU(2)-structure to obtain more detailed classes.
\subsubsection{Simple ansatz: warped SU$(2)$-holonomy}\label{subsec: CaseA1}
The easiest way to solve \eqref{SU2susy2}-\eqref{1casefluxcondition} is  to impose that the part of the metric orthogonal to $v$ has warped SU$(2)$-holonomy, i.e.
\beq
e^{\Delta}\omega_2=\tilde\omega_2= e^{\Delta}j_2=\tilde j_2,
\eeq
with $\tilde \omega_2,~\tilde j_2$ closed.
The local form of the solution is then
\begin{align}
ds^2&= H^{-2/3} ds^2(\mathbb{R}_{1,3})+ H^{1/3}\bigg( ds^2(M_4)+H\big(d\rho^2+ \rho^2 ds^2(S^2)\big)\bigg),\nn\\[2mm]
e^{2C}F_2&= c \tilde j_2,~~~H=1+\frac{c}{\rho}.
\end{align}
with $M_4$ any SU$(2)$-manifold. These solutions are all non compact and the warp factor is not indicative of a simple brane set up. Following \cite{Jarv:2000zv}, this could be a KK6-M2 system or something more exotic like the lift of a O6-D2 system.

\subsubsection{Squashed $S^3$ ansatz}\label{subsec: CaseA2}
Let's consider the simple case where the SU(2)-structure is defined on a squashed $S^3$ trivially fibered over an interval $y$. Here we take the vielbein to be
\begin{equation}
v= -e^{-2 \Delta} d \rho , \qquad w = \frac{e^{C_1}}{2} (\omega_1+ i \omega_2) , \qquad u = \frac{e^{C_2}}{2} \omega_3 + i e^{K-2\Delta+h(y)} d y,
\end{equation}
where  $C_1,C_2,K,\Delta$ are warp factors which depend on $\rho = e^{2\Delta +C}$ and $y$ only, $h$ is an arbitrary function of $y$ and $\omega_i$ are the right-invariant one-form defined on $S^3$, which satisfies
\begin{equation}
d \omega_i = - \frac{1}{2}\eps_{ijk} \omega^j \wedge \omega^k. 
\end{equation}
Notice that this means we are enhancing the R-symmetry to SU$(2)\times \text{U}(1)$ as $(\omega_1+ i \omega_2)$ comes with a phase $e^{i\psi}$, where $\partial_{\psi}$ is the U(1) of the Hopf fibration. From \eqref{SU2susy2} we get the following three equations:
\begin{equation}
\partial_\rho e^{C_1+C_2+\Delta} = \partial_\rho e^{C_1+K-\Delta} = 0 , \qquad \partial_y  e^{C_1+C_2+\Delta} = 2 e^{C_1+K-\Delta+h}.
\end{equation}
These are solved by defining $C_1$ and $C_2$ in terms of two functions $f(y)$ and $g(y)$ such that
\begin{equation}
e^{C_1} = e^{\Delta-K-h} \frac{f}{g} ~~~~e^{C_2} = e^{K-2\Delta+h} g,~~~~g = \frac{2 f}{f'} .
\end{equation}
At this point, because it simplifies later expressions, we choose the arbitrary function $h$ and use diffeomorphism invariance in y to fix $f$ as
\beq
e^{h}= f',\qquad  f= 2 \sqrt{y},
\eeq
without loss of generality. This fixes the metric as
\beq\label{eq:mansmet}
ds^2 = e^{2\Delta} ds^2(\text{Mink}_4) + e^{-4\Delta}\bigg(d\rho^2+   \rho^2 ds^2(S^2)+\frac{e^{2K}}{y}dy^2+ 4 y e^{2K}(d\psi+ \eta)\bigg)+ \frac{e^{2\Delta-2K}}{4}ds^2(\tilde{S}^2)
\eeq
where $\eta$ is a potential for the Kahler form on $\tilde{S}^2$.
The two conditions involving the flux in \eqref{1casefluxcondition} impose that
\begin{equation}
e^{-6\Delta+2K}= \frac{1}{8} \partial_y (e^{-2K}),  
\end{equation}
which allows one to define $\Delta$ in terms of K. What remains of \eqref{1casefluxcondition} just defines the flux $F_2$, which can be used to express the M-theory 4-form as
\beq
G = \frac{\rho^2}{4} \bigg(\partial_{\rho}(e^{-2K}) \omega_1\wedge \omega_2+ \partial_{\rho y}^2 (e^{-2K}) \omega_3 \wedge d y - \frac{y}{2} \partial_y^2 (e^{-4K}) \omega_3 \wedge d \rho\bigg)\wedge \text{Vol}(S^2)\nn.
\eeq
All we need to do now is  solve the Bianchi identity of the flux, which away from localised sources leads to just one partial differential equation (PDE) in $e^{K}$, namely 
\beq\label{eq:squshedS3PDE}
\frac{1}{\rho^2}\partial_{\rho}(\rho^2\partial_{\rho}(e^{-2K}))+ \frac{y}{2} \partial_{y}^2(e^{-4K})=0.
\eeq
This bares a striking similarity to the PDE governing the intersecting D8-D6-NS5  brane system of \cite{Imamura:2001cr} - the only difference is the $y$ factor. We also find deformations of this PDE in section \ref{sub:massiveIIA} of the appendix. Of course here this is only a formal similarity and the class of solutions currently under investigation is unlikely to have any relation to a Mink$_6$ class in massive IIA. At any rate, the form of the metric in \eqref{eq:mansmet}, and a relatively simple PDE governing the physical fields, makes this class appear promising for finding compact solutions in M-theory\footnote{One might also wonder if AdS solutions exist here. We checked that all of the SUSY constraint for the most general AdS$_5$ solution one can embed in this class can be solved - but the flux always has a leg in the AdS direction breaking the SO$(2,4)$ symmetry; moreover, the Bianchi identity is not satisfied.}

\subsection{Case B: $\alpha \neq 0$, the M5-brane}\label{sec: CaseB}
In this case $\alpha$ is non zero, however one can see that it actually factors out from all of the supersymmetry conditions and in the end nothing physical depends on its specific value, as long as $\sin\alpha \neq 0$ (in which case we fall into the class of the previous section). The 1-form supersymmetry constraints are
\begin{equation}
d(e^{2 \Delta + C}) + e^{2\Delta} v = d(e^{2 \Delta} w) = d(e^{- \Delta} u) = 0
\end{equation}
which we can be solved without loss of generality by introducing local coordinates $x_1,x_2,x_3,x_4$ and
\beq
\rho = e^{2\Delta+C},
\eeq
such that the five-dimensional vielbein becomes
\begin{equation}\label{eq: veilbein}
v = - e^{- 2 \Delta} d \rho , \quad w = e^{- 2 \Delta} (d x_1 + i d x_2) , \quad u = e^{\Delta} (d x_3 + i d x_4) , \quad \rho = e^{2 \Delta + C},
\end{equation}
which span an identity-structure - so we have completely local expressions for solutions in this class.
The rest of the supersymmetry conditions, that are not implied by the one-forms, involve the flux, namely 
\begin{align}
&d(e^{4 \Delta} u_1 \wedge u_2 ) + e^{4 \Delta} \star_5 F_2 = 0,\label{2casefluxconditiona}\\[2mm]
&d(e^{2C + 2 \Delta} w_2 \wedge v) + e^{2C + 2 \Delta} w_1 \wedge F_2 = 0,\label{2casefluxconditionb}\\[2mm]
&\cos\alpha(d(e^{-2\Delta}( w_1\wedge w_2)+ e^{-2\Delta}F_2\wedge v)=0.\label{2casefluxconditionc}
\end{align}
The definition of $F_2$ can be read from \eqref{2casefluxconditiona} where we can make  use of \eqref{eq: veilbein} to take the Hodge dual:
\beq
F_2=- \rho^2 e^{-2C} \left(  \partial_{\rho} e^{-6 \Delta} d x_1 \wedge d x_2+ \partial_{x_2} e^{-6 \Delta} d \rho \wedge d x_1 + \partial_{x_1} e^{-6 \Delta} d x_2 \wedge d \rho  \right).
\eeq 
We then plug this back into \eqref{2casefluxconditionb} which imposes
\beq
\partial_{x_3}e^{2\Delta}=\partial_{x_4}e^{2\Delta}=0,
\eeq
so that $\partial_{x_3}$, $\partial_{x_4}$ are necessarily isometries of the solution - this automatically solves \eqref{2casefluxconditionc} without imposing any further restriction on $\alpha$, then nothing physical depends on it. Thus the local form of all solutions in this class is
\begin{align}\label{eq:M5metflux}
ds^2&= e^{2\Delta}ds^2(\text{Mink}_{6})+ e^{-4\Delta}\big(dx_1^2+ dx_2^2+d\rho^2+ \rho^2 ds^2(S^2)\big),\\[2mm] 
G &=- \rho^2  \left(  \partial_{\rho} e^{-6 \Delta} d x_1 \wedge d x_2+ \partial_{x_2} e^{-6 \Delta} d \rho \wedge d x_1 + \partial_{x_1} e^{-6 \Delta} d x_2 \wedge d \rho  \right)\wedge \text{Vol}(S^2),
\end{align}
where the Bianchi identity of $G$ imposes that the warp factor obeys
\begin{equation}
\frac{1}{\rho^2} \partial_{\rho}\left( \rho^2 \partial_{\rho} e^{-6\Delta} \right)+ \partial_{x_1}^2 e^{-6\Delta} + \partial_{x_2}^2 e^{-6\Delta}  = 0 \,
\end{equation}
away from localised sources. This is a five-dimensional Laplace equation expressed in coordinates that make the the $SO(3)$ symmetry of it's solutions manifest - thus this entire class of solutions is nothing more than  M5-branes with some rotational symmetry in their co-dimensions. It is easy to check that the supersymmetry conditions \eqref{eq: veilbein}-\eqref{2casefluxconditionc} are actually compatible with the supersymmetry conditions \eqref{SU2susy1}-\eqref{1casefluxcondition} of case A, which means that the physical solution of Case B can actually be embedded in Case A by taking \eqref{eq: veilbein} to define the SU(2)-structure as in \eqref{eq:SU2veil}.

We shall see in the next section that all solutions in type II which have internal Killing spinors with non equal norm, descend via dimensional reduction and T-duality, from this class.

\section{Classes of solution in type II with non equal spinor norm}\label{sec:nonequalnorm}
Supersymmetric type II solutions with Mink$_4\times S^2$ factors were partially classified in \cite{Macpherson:2016xwk,Apruzzi:2018cvq}. Similar to the M-theory classification of section \ref{sec:Mtheoryclassification}, the starting point is a decomposition of the ten-dimensional Majorana--Weyl Killing spinors of the form
\beq\label{eq:spinordef}
\epsilon_1= \zeta_+ \otimes (\xi\otimes \tilde {\eta}^1+\sigma_3\xi\otimes\hat \gamma \tilde {\eta}^1)+\text{m.c.},~~~\epsilon_2= \zeta_+ \otimes (\xi\otimes \tilde {\eta}^2\pm\sigma_3\xi\otimes\hat \gamma \tilde {\eta}^2)+\text{m.c.},
\eeq
where $\pm$ is taken in IIB/IIA. Once more $\xi$ is a Killing spinor on $S^2$ obeying \eqref{eq:S2KSE} and now $\eta^i$ defines a non-chiral spinor in four dimensions. The metric decomposes as
\beq\label{eq:tpeIImetric}
ds^2= e^{2A}ds^2(\text{Mink}_4)+ e^{2C}ds^2(S^2) +ds^2(\text{M}_4), 
\eeq
where $e^{2A}$ and $e^{2C}$ depend on the coordinates on M$_4$ only.
However, unlike the M-theory cases, the classifications in ten dimensions \cite{Macpherson:2016xwk,Apruzzi:2018cvq} are not completely general, as we will now explain. A consequence of supersymmetry is that
\beq
d(e^{\mp A}(|\eta_1|^2\pm|\eta_2|^2))=0,
\eeq
which means we can define constants $c_{\pm}$ such that
\beq
|\eta_1|^2\pm|\eta_2|^2 = c_{\pm} e^{\pm A},
\eeq
$c_+$ can be tuned to any positive constant without loss of generality, but tuning $c_-$ can have marked physical effects - namely the physics of solutions with generic $c_-$ is quite different to the physics when $c_-=0$, i.e. the case of equal spinor norm. In \cite{Macpherson:2016xwk,Apruzzi:2018cvq} it is only the equal norm case that is classified, so in this section we shall study the possible solutions in type II for generic $c_-$, i.e. non equal spinor norm. As we shall see, it will turn out that such solutions always exhibit two uncharged U(1) isometries, so all type IIB solutions are contained in the classification of type IIA up to a T-duality, and further, all type IIA solutions have the Romans mass set to zero - so descend from our M-theory classification.

In the next section we shall study the unequal norm cases in type IIA.

\subsection{Non-equal norm in IIA}\label{sec:noneqnormIIA}
We begin our analysis in IIA, our goal here is not to give a detailed classification, rather we seek to show that all solutions in this class descend from M-theory. This statement turns out to actually be true for any warped Mink$_4$ solution - so we will not actually assume an $S^2$ factor here, merely a decomposition of the metric of the form
\beq
ds^2= e^{2A} ds^2(\text{Mink}_4)+ ds^2(\text{M}_6),
\eeq
and likewise for the fluxes, with all physical feilds supported by M$_6$ only.

In the conventions of \cite{DeLuca:2018buk}, supersymmetry depends on the existence of a non-chiral IIA spinor on warped Mink$_4\times M_6$ that takes the form
\beq\label{eq:IIA10dsinpor}
\epsilon = \zeta_+\otimes \chi+ \zeta^c_+\otimes \chi^c
\eeq
where $\zeta_+$ is a positive chirality spinor in four dimensions, $\chi$ a non-chiral six-dimensional spinor and the $c$ superscript labels Majorana conjugation\footnote{One can without loss of generality take $B^{(10)}=\mathbb{I}\otimes B^{(6)}$ where $B^{(6)}B^{(6)*}=\mathbb{I}$ as the relevant intertwiner so that $\zeta^c_+=\zeta^*_+$ and $\chi^c=B^{(6)}\chi^*$.} such that $\epsilon^c=\epsilon$. Supersymmetry is then implied by the following spinorial conditions on $M_6$
\begin{subequations}
\begin{align}
&\bigg[\partial A+\frac{e^{\Phi}}{4}\big(F_0+F_2\hat\gamma^{(6)}+F_4+i \star_6 F_6\big)\bigg]\chi=0,\label{eq:6dIIASUSYcond1}\\[2mm]
&\bigg[\partial \Phi+\frac{1}{2}H\hat\gamma^{(6)}+\frac{e^{\Phi}}{4}\big(5F_0+3F_2\hat\gamma^{(6)}+F_4-i \star_6 F_6\big)\bigg]\chi=0,\label{eq:6dIIASUSYcond2}\\[2mm]
&\bigg[\nabla_a +\frac{1}{4}H_a\hat \gamma^{(6)}+\frac{e^{\Phi}}{8}\big(F_0-F_2\hat\gamma^{(6)}+F_4-i \star_6 F_6\big)\gamma^{(6)}_a\bigg]\chi=0,\label{eq:6dIIASUSYcond3}
\end{align}
\end{subequations}
where $\Phi$ is the dilaton, $\hat\gamma^{(6)}$ the chiral operator and where in \eqref{eq:6dIIASUSYcond3} we correct a typo appearing in \cite{DeLuca:2018buk}.
Here $F_n$ labels the RR flux magnetically coupled to a space-filling brane
\beq
F_{\text{RR}}= F+e^{4A}\text{Vol}_4\wedge \star_6 \lambda(F)
\eeq
however, as we shall see shortly, the $0,2,4$ forms will turn out to be purely magnetic so we need not worry about this distinction for long.
Using \eqref{eq:6dIIASUSYcond1} and \eqref{eq:6dIIASUSYcond3} it is possible to establish that $d(e^{-A}|\chi|^2)=d(e^{A}\chi^{\dag}\hat\gamma\chi)=0$, while  \eqref{eq:6dIIASUSYcond2} implies that $d(e^{2A-\Phi}\overline{\chi}\chi)=0$,  which one can use to fix
\beq\label{eq:norm}
|\chi|^2= c_+e^{A},~~~~\chi^{\dag}\hat\gamma^{(6)}\chi= c_- e^{-A},~~~~\overline{\chi}\chi= c e^{-2A+\Phi},~~~~\overline{\chi}\hat\gamma^{(6)}\chi=0,
\eeq
where the final condition follows in general whenever $(\hat\gamma^{(6)})^T=-\hat\gamma^{(6)}$.
$c_{\pm},c$ are constants and we set $c_+=2$ without loss of generality, $c_-=0$ is the cases of equal internal spinor norms and $c$ parameterises the G-structure of the internal 6 manifold; in particular, $c=0$ corresponds to an orthogonal SU$(2)$-structure.  Using the relations
\beq
(\chi^{\dag}\gamma^{(6)}_{a_1,...a_n}\chi)^*=(-1)^{\frac{n}{2}(n-1)}\chi^{\dag}\gamma^{(6)}_{a_1,...a_n}\chi,~~~~~(\chi^{\dag}\gamma^{(6)}_{a_1,...a_n}\hat\gamma^{(6)}\chi)^*=(-1)^{\frac{n}{2}(n+1)}\chi^{\dag}\gamma^{(6)}_{a_1,...a_n}\hat\gamma^{(6)}\chi,
\eeq
which actually hold for Hermitian gamma matrices in any dimension, we find that the rank 2,3,6 forms which come from the bi-linear $\chi \otimes \chi^{\dag}$ and the rank 1,2,5,6 forms constructed from $\hat\gamma^{(6)}\chi \otimes \chi^{\dag}$  are purely imaginary, while the rest are real. Using this fact, it is possible to show from the inner-products of \eqref{eq:6dIIASUSYcond1}-\eqref{eq:6dIIASUSYcond2} with $\chi$ and $\hat\gamma^{(6)}\chi$ that
\beq\label{eq: F0F6rules}
\star_6 F_6= F_0 \chi^{\dag}\hat\gamma^{(6)}\chi=0,
\eeq
which come from the imaginary and real parts of the inner products respectively.
Equal spinor norm is equivalent to $\chi^{\dag}\hat\gamma^{(6)}\chi=0$ in this language, so it is clear that a non vanishing Romans mass is only possible for equal spinor norm. As such all type IIA solutions with non equal spinor norm have $F_0=0$ so can be lifted to M-theory.

As we shall show later in section \ref{sec:connectiontoM5}, the Mink$_4\times S^2$ non equal norm solutions in IIA necessarily descend from section \ref{sec: CaseB} by dimensional reduction. % For the sake of consistency however, we study the spinors of the Mink$_4\times S^2$ system in IIA in Appendix \ref{sec:noneqnormIIAmore} where we show that in addition to $F_0=0$, non equal spinor norm always imposes an uncharged $\text{U}(1)\times \text{U}(1)$ isometry in the internal space orthogonal to Mink$_4\times S^2$.
In the next section we turn our attention to type IIB solutions with non equal norm, as we shall see, this requires a more detailed analysis.

\subsection{Non equal norm in IIB}
In this section we study the Killing spinor conditions in IIB. Our goal is to establish that when the spinor norms are not equal, there are always two flavour \text{U}(1) isometries - however, most of the results in this  section apply irrespective of the particular values of the norms.

In IIB we shall define our ten-dimensional Killing spinors as in \ref{eq:spinordef}, but redefine the 4d spinors as
\beq
\tilde\eta^1=\frac{1}{2}(\eta^1+\eta^2),\quad \tilde\eta^2=\frac{1}{2}(\eta^1-\eta^2).
\eeq
As we establish in Appendix \ref{sec:AppIIB}, supersymmetry imposes the following set of four-dimensional spinoral constraints
\begin{subequations}
\begin{align}
&\partial A\eta^1+\frac{e^{\Phi}}{4}\bigg[-(f_1\hat\gamma + i g_3)\eta^2+ (f_3 \hat\gamma+i g_1)\eta^1\bigg]=0,\label{eq:susyIIB4d1}\\[2mm]
&\partial A\eta^2-\frac{e^{\Phi}}{4}\bigg[-(f_1\hat\gamma + i g_3)\eta^1+ (f_3 \hat\gamma+i g_1)\eta^2\bigg]=0,\label{eq:susyIIB4d2}\\[2mm]
&\partial\Phi\eta_1-\frac{1}{2}(H_3+i H_1\hat\gamma)\eta^2+e^{\Phi}\bigg[-f_1\hat\gamma\eta^2+\frac{1}{2}(f_3\hat\gamma+ig_1)\eta^1\bigg]=0\label{eq:susyIIB4d3},\\[2mm]
&\partial\Phi\eta_2-\frac{1}{2}(H_3+i H_1\hat\gamma)\eta^1-e^{\Phi}\bigg[-f_1\hat\gamma\eta^1+\frac{1}{2}(f_3\hat\gamma+ig_1)\eta^2\bigg]=0\label{eq:susyIIB4d4},\\[2mm]
&\partial C \eta^1-i e^{-C}\hat\gamma\eta^2-\frac{i}{2}H_1\hat\gamma \eta^2+\frac{e^{\Phi}}{4}\bigg[-(f_1\hat\gamma-i g_3)\eta^2+(f_3\hat\gamma-i g_1)\eta^1\bigg]=0\label{eq:susyIIB4d5},\\[2mm]
&\partial C \eta^2-i e^{-C}\hat\gamma\eta^1-\frac{i}{2}H_1\hat\gamma \eta^1-\frac{e^{\Phi}}{4}\bigg[-(f_1\hat\gamma-i g_3)\eta^1+(f_3\hat\gamma-i g_1)\eta^2\bigg]=0\label{eq:susyIIB4d6},\\[2mm]
&\nabla_a\eta_1-\frac{1}{4}\bigg((H_3)_a+ i (H_1)_a \hat\gamma\bigg)\eta^2+\frac{e^{\Phi}}{8}\bigg[-(f_1\hat\gamma-i g_3)\gamma_a\eta_2+(f_3\hat\gamma-i g_1)\gamma_a\eta_1\bigg]=0\label{eq:susyIIB4d7},\\[2mm]
&\nabla_a\eta_2-\frac{1}{4}\bigg((H_3)_a+ i (H_1)_a \hat\gamma\bigg)\eta^1-\frac{e^{\Phi}}{8}\bigg[-(f_1\hat\gamma-i g_3)\gamma_a\eta_1+(f_3\hat\gamma-i g_1)\gamma_a\eta_2\bigg]=0,\label{eq:susyIIB4d8}
\end{align}
\end{subequations}
where $f,g,H_1,H_3$ are forms related to the NS and RR fluxes defined as in Appendix \ref{sec:AppIIB} while $\Phi,C,A$ are function on $M_4$.

From the six-dimensional zero-form conditions \eqref{eq:normIIB}, imposing that the physical quantities are SU(2) singlets leads to the following constaints for the four-dimensional spinors:
\begin{align}\label{eq:IIBbizero1}
&|\eta^1|^2=|\eta^2|^2=e^{A},~~~~\eta^{1\dag}\hat\gamma \eta^2+\eta^{2\dag}\hat\gamma \eta^1=0,\nn\\[2mm]
&\eta^{1\dag}\eta^2=\eta^{2\dag}\eta^1= c_- e^{-A},~~~~\eta^{1\dag}\hat\gamma \eta^1+\eta^{2\dag}\hat\gamma \eta^2=0,
\end{align}
Working a bit harder one can establish another scalar condition from the difference of\\$\overline{\eta^{2}}(2$\eqref{eq:susyIIB4d1}+2\eqref{eq:susyIIB4d5}-\eqref{eq:susyIIB4d3}) and $\overline{\eta^{1}}(2$\eqref{eq:susyIIB4d2}+2\eqref{eq:susyIIB4d6}-\eqref{eq:susyIIB4d4}) \footnote{We define $\overline{\eta}= (\eta^c)^\dag$ where $\eta^c = B \eta^*$ and the conventions for $B$ are given in Appendix \ref{sec:AppIIB}.}, namely
\begin{equation}
\overline{\eta^2} \hat \gamma \eta^1 = 0.
\end{equation}
One can  solve all of the scalar conditions with the following decomposition of $\eta^i$ in terms of a single spinor $\eta$, and complex functions $a,b$
\beq\label{eq:spinordecompostionIIB}
e^{-\frac{1}{2}A}\eta^1= \sin\left(\frac{\alpha}{2}\right)\eta +\cos\left(\frac{\alpha}{2}\right)\big(a\hat\gamma\eta+ b\eta^c\big),~~~~e^{-\frac{1}{2}A}\eta^2= \sin\left(\frac{\alpha}{2}\right)\eta -\cos\left(\frac{\alpha}{2}\right)\big(a\hat\gamma\eta+ b\eta^c\big)
\eeq 
where
\beq
|\eta|^2=1,~~~ \eta^{\dag}\hat\gamma \eta = 0,~~~ |a|^2+ |b|^2=1,\quad e^{2A} \cos\alpha=-c_-,
\eeq
so that $\alpha$ parametrises the difference in norms of the spinors, which are equal only when $\cos\alpha=0$.

As shown in \cite{Apruzzi:2014qva,Macpherson:2017mvu}, the spinor $\eta$ is sufficient to define an identity-structure on $M_4$, which is expressed in terms of the vielbein $\{v_1,v_2,w_1,w_2\}$ 
\beq\label{eq:M4vielbeinIIB}
\eta^{\dag}\gamma_a \eta= v_1,~~~~\eta^{\dag}\gamma_a\hat\gamma \eta= i v_2,~~~~\overline{\eta}\gamma_a\hat\gamma \eta= w= w_1+ i w_2.
\eeq  
The definition of $w$  is such that $\eta^c=\frac{1}{2} \overline{w}\hat{\gamma} \eta$ , which we can use to fix the phase of $b$ in \eqref{eq:spinordecompostionIIB} by rotating the vector $w$, so we choose $b$ to be real and expand $a$ as
\beq
b= |b|,\quad a=a_1+i a_2 .
\eeq
Let us now consider the vector
\beq\label{eq: IIBisometry}
k=\overline{\eta^2}\gamma^a\hat\gamma\eta^1 \partial_a,
\eeq
which can be rewritten as a one form in terms of the vielbein \eqref{eq:M4vielbeinIIB}
\beq\label{eq:Killing1-formIIB}
e^{-A}k=a\big(1+\cos\alpha\big)\big(b v_1+(a_1 w_1- a_2 w_2)\big)- \big(\cos\alpha w_1- i w_2\big).
\eeq
We will spend the bulk of this  section  proving that $k$ is a Killing vector parameterising a $\text{U}(1)\times \text{U}(1)$  flavour symmetry in the solutions. However let us first perform some  consistency checks: we notice that for generic values of $\alpha$ both the real and imaginary components of $k$ cannot vanish, so $k$  necessarily defines 2 real and non zero vectors. The exception is when $\cos\alpha=0$, which is the case of equal  spinor norm studied in \cite{Apruzzi:2018cvq} - here the number of isometries we find depends on the values of $a_1,a_2,b$. In particular there is the following reduced isometry structure when the functions of the spinor ansatz are tuned as follows
\begin{align}
&\text{0 isometries:}~~~ \cos\alpha=a_1=b=0,\nn\\[2mm]
&\text{1 isometries:}~~~ \cos\alpha=a_1=0,~b\neq 0\nn,
\end{align}
which can happen because $k$ is respectively zero or purely imaginary in  these two cases only - so $k$ is consistent with the isometry structure of the known classes with equal spinor norm, a strictly non vanishing for non equal norm, a promising sign we are on the correct track.

The first thing to establish is that $k$ is an isometry with respect to scalar physical fields. To do so one can take the sum of the inner products of \eqref{eq:susyIIB4d1}, \eqref{eq:susyIIB4d2} with $\hat\gamma\eta^2$, $\hat\gamma\eta^1$ respectively and repeat this operation  for  \eqref{eq:susyIIB4d3},\eqref{eq:susyIIB4d4} and \eqref{eq:susyIIB4d5},\eqref{eq:susyIIB4d6}. This leads to the conditions
\begin{equation}
k^a \partial_a A = k^a \partial_a \Phi = k^a \partial_a C = 0
\end{equation}
which means that $A,C,\Phi$ do not depend on the th$k$ directions.

Now let's prove that $k$ is Killing with respect to the metric, i.e., that $\nabla_{(a} k_{b)}=0$. A straightforward calculation leads to:
\begin{equation}
\nabla_a k_b = \frac{1}{4} H_{3 \, abd} (\overline{\eta^{1}} \gamma^d \hat \gamma \eta^1+\overline{\eta^{2}} \gamma^d \hat \gamma \eta^2 )+ \frac{e^{\Phi}}{8} (\star f_1+i g_3)_{abd}(\overline{\eta^{1}} \gamma^d \hat \gamma \eta^1-\overline{\eta^{2}} \gamma^d \hat \gamma \eta^2) + \frac{e^\Phi}{4} (f_3+ i \star g_1)_{abd}\overline{\eta^{2}} \gamma^d \eta^1,\nn
\end{equation}
and by the manifest anti-symmetry of the right hand side, it is immediate to notice that $k$ is a Killing vector. This expression can be further simplified, indeed, using the combinations of equations (2\eqref{eq:susyIIB4d1}+\eqref{eq:susyIIB4d5}-\eqref{eq:susyIIB4d3}) and (2\eqref{eq:susyIIB4d2}+\eqref{eq:susyIIB4d6}-\eqref{eq:susyIIB4d4}), which read
\begin{align}
&\partial (2A+C-\Phi) \eta^1+\frac{1}{2}H_3\eta^2-i e^{-C}\hat\gamma\eta^2+\frac{e^{\Phi}}{4}\bigg[-(-f_1\hat\gamma+i g_3)\eta^2+(f_3\hat\gamma-i g_1)\eta^1\bigg]=0\label{eq:susyIIB4mod1},\\[2mm]
&\partial (2A+C-\Phi) \eta^2+\frac{1}{2}H_3\eta^1-i e^{-C}\hat\gamma\eta^1-\frac{e^{\Phi}}{4}\bigg[-(-f_1\hat\gamma+i g_3)\eta^1+(f_3\hat\gamma-i g_1)\eta^2\bigg]=0\label{eq:susyIIB4mod2}, 
\end{align}
one may establish  from the sum of $\overline{\eta^{2}} \gamma_{ab} \hat \gamma$\eqref{eq:susyIIB4mod1} and $\overline{\eta^{1}} \gamma_{ab} \hat \gamma$\eqref{eq:susyIIB4mod2}, that
\begin{align}
&\frac{1}{4} H_{3 \, abd} (\overline{\eta^{1}} \gamma^d \hat \gamma \eta^1+\overline{\eta^{2}} \gamma^d \hat \gamma \eta^2 )+ \frac{e^{\Phi}}{8} (\star f_1+i g_3)_{abd}(\overline{\eta^{1}} \gamma^d \hat \gamma \eta^1-\overline{\eta^{2}} \gamma^d \hat \gamma \eta^2) + \frac{e^\Phi}{4} (f_3+ i \star g_1)_{abd}\overline{\eta^{2}} \gamma^d \eta^1\nn\\
&=-2 \partial_{[a}(2A+C-\Phi)k_{b]} + \frac{i}{2} e^{-C}(\overline{\eta^1}\gamma_{ab}\eta^1 + \overline{\eta^2}\gamma_{ab}\eta^2).
\end{align}
We therefore arrive at
\begin{equation}
\label{eq:nablakIIB}
\nabla_a k_b =-2 \partial_{[a}(2A+C-\Phi)k_{b]} + \frac{i}{2} e^{-C}(\overline{\eta^1}\gamma_{ab}\eta^1 + \overline{\eta^2}\gamma_{ab}\eta^2).
\end{equation}
This expression is particularly useful because allows us to establish that the isometry group of $k$ is $\text{U}(1)\times \text{U}(1)$. To prove this  we need to show that the Lie bracket $[k, \overline{k}]=0$  which is equivalent to
\beq
\overline{k}^a \nabla_a k_b\in\mathbb{R}.
\eeq
Given our spinor decomposition and the definition of the four-dimensional vielbein in terms of a single spinor $\eta$, one can show that
\beq
(\overline{\eta^1}\gamma_{ab}\eta^1 + \overline{\eta^2}\gamma_{ab}\eta^2) = -4i e^{A}k_{[a} (v_2)_{b]}.
\eeq
It is then a simple matter to calculate 
\beq
\overline{k}^a \nabla_a k_b = -|k|^2\big(\partial_{b} (2A-\Phi+C) - e^{A-C}(v_2)_b\big),
\eeq
which is clearly real - so the isometry group of $k$ is indeed $\text{U}(1)\times \text{U}(1)$.

The next task is to prove that $\text{U}(1)\times \text{U}(1)$ is a flavour symmetry - which means that the spinors $\eta^{1,2}$ are not charged under these isometries. This statement can be caste as the vanishing of the spinoral Lie derivative along $k$, i.e.
\begin{equation}
\label{eq:spinorial_Lie}
\mathcal{L}_k \eta^i = k^a \nabla_a \eta^i + \frac{1}{4} \nabla_a k_b \gamma^{ab} \eta^i = 0 .
\end{equation} 
To prove this we will need the following algebraic relations, which can be demonstrated using standard Fierz identity techniques:
\begin{equation}
\label{eq:algebraic_spinorial}
k \eta^1 = (\overline{\eta^2} \eta^1) \hat \gamma \eta^1 , \qquad k \eta^2 = - (\overline{\eta^2} \eta^1) \hat \gamma \eta^2 , \qquad i e^A v_2 \eta^2 = \hat \gamma \eta^1, \qquad i e^A v_2 \eta^1 = \hat \gamma \eta^2 ,
\end{equation}
where $k$ and $v_2$ must now be understood as gamma matrices (via Clifford map).
Let us fix $i = 1$ (the discussion for $i = 2$ is specular) and start by evaluating the first term in \eqref{eq:spinorial_Lie}. Contracting \eqref{eq:susyIIB4d7} with $k^a$ and using the first two relations of \eqref{eq:algebraic_spinorial} one finds:
\begin{equation}
k^a \left(\nabla_a\eta^1-\frac{1}{4}\big((H_3)_a+ i (H_1)_a \hat\gamma\big)\eta^2 \right)+\frac{e^{\phi}}{8}(\overline{\eta^2} \eta^1)\left(-(-f_1\hat\gamma+i g_3)\hat\gamma\eta^2+(f_3\hat\gamma-i g_1)\hat\gamma\eta^1\right)=0 .
\end{equation}
Now we can substitute for the RR fluxes using \eqref{eq:susyIIB4mod1} multiplied by $\hat\gamma$. With some manipulations one realises
\begin{equation}
k^a \left(\nabla_a\eta_1-\frac{1}{4} (\star H_3+ i H_1)_a \hat\gamma\eta^2 \right) - \frac{1}{2}(\overline{\eta^2} \eta^1) \left(\partial (2A+C-\Phi) \hat\gamma\eta^1+i e^{-C}\eta^2 \right) = 0.
\end{equation}
The term $k^a(\star H_3+ i H_1)_a$ can be evaluated from the inner product of \eqref{eq:susyIIB4d3} with $\eta^1$ and \eqref{eq:susyIIB4d4} with $\eta^2$, indeed this leads to
\begin{equation}
\label{eq:IIBcontraction1}
k^a(\star H_3+ i H_1)_a=k^a (f_{1})_a = 0,
\end{equation}
note that this implies that when the norms are not equal, so that $k$ defines two non vanishing Killing vectors in general,  $H_1$ has no legs in the isometry directions, while $H_3$ is parallel to both. We are left with the expression
\begin{equation}
\label{eq:knabla_etaIIB}
k^a \nabla_a\eta_1 = \frac{1}{2}(\overline{\eta^2} \eta^1) \left(\partial (2A+C-\Phi) \hat\gamma\eta^1+i e^{-C}\eta^2 \right) = \frac{1}{2} \left(\partial (2A+C-\Phi) k\eta^1-i e^{-C} \hat \gamma k \eta^2 \right).
\end{equation}
Now we consider the second term in \eqref{eq:spinorial_Lie}: using \eqref{eq:nablakIIB} one may write
\begin{equation}
(\nabla_a k_b ) \gamma^{ab} \eta^1 = -2(\partial_a(2A+C-\phi) k_b - e^{A-C} k_{a}(v_2)_b) \gamma^{ab} \eta^1 .
\end{equation}
Since $k_a$ is orthogonal to $\partial_a(2A+C-\phi)$ and $(v_2)_a$, we can take $\gamma^{ab} = \gamma^a \gamma^b$ in the previous expression and, using \eqref{eq:algebraic_spinorial} we find 
\begin{equation}
\label{eq:killing_etaIIB}
(\nabla_a k_b) \gamma^{ab} \eta^1 = - 2\partial(2A+C-\phi) k \eta^1 + 2 i e^{-C} \hat \gamma k \eta^2 .
\end{equation}
Now one simply combines the contributions form \eqref{eq:knabla_etaIIB} and \eqref{eq:killing_etaIIB} to establish that
\begin{equation}
\mathcal{L}_k \eta^1 = 0.
\end{equation}
It is easy to perform the same steps for $\mathcal{L}_k\eta^2$ and we  therefore find that $\text{U}(1)\times \text{U}(1)$ is a flavour symmetry. 

The last step is to prove that also the fluxes are not charged under the isometries. Extrapolating from the classification of \cite{Grana:2005sn}, we know that the RR fluxes are defined in terms of bi-linears of $\eta^i$ and the NSNS 3-form. Since the spinors are uncharged, the same is true of the bi-linears and so we need only establish that the NSNS 3-form is a $\text{U}(1)\times \text{U}(1)$ singlet. We begin by considering  $\overline{\eta^{1}}$\eqref{eq:susyIIB4d1}, $\overline{\eta^{1}}$\eqref{eq:susyIIB4d3} + $\overline{\eta^{2}}$\eqref{eq:susyIIB4d4}, $\overline{\eta^{2}} \hat \gamma$\eqref{eq:susyIIB4d1}, $\overline{\eta^{2}} \hat \gamma$\eqref{eq:susyIIB4d5}; these, together with \eqref{eq:IIBcontraction1}, allow us to find the following relations:
\begin{equation}
\label{eq:IIBcontraction2}
k^a (H_1)_a = k^a (\star H_3)_a = k^a (f_1)_a = k^a (\star f_3)_a = k^a (g_1)_a = k^a (\star g_3)_a = 0 .
\end{equation}    
The first, with the Bianchi identity $d(e^{2C}H_1)=0$ implies $\mathcal{L}_k (e^{2C}H_1) = 0$. Now let's consider the equation of motions for the B-field,
%\begin{equation}
%(e^{-2\Phi} \star H) + F_1 \wedge F_7 + F_5 \wedge F_3 = 0 ;
%\end{equation}
this implies the following equation
\begin{equation}
d(e^{4A + 2C - 2 \Phi} \star H_3) = e^{4A+2C} (f_1 \wedge \star f_3 + \star g_3 \wedge g_1).
\end{equation}
Contracting this with $k$ and using \eqref{eq:IIBcontraction2} we get that
\begin{equation}
\imath_k d(\star H_3) + d(\imath_k \star H_3) = \mathcal{L}_k (\star H_3) = 0,
\end{equation}
which proves that $H_1$ and $H_3$ are singlets under $\text{U}(1)\times \text{U}(1)$.

We have now established that all Mink$_4\times S^2$ solutions in type IIB with non equal spinor norm necessarily have a $\text{U}(1)\times \text{U}(1)$ flavour symmetry. In the next section we shall show that they all actually follow from IIA via T-duality and can then be lifted to class B in M-theory.

\subsection{Relation to the M5-brane}\label{sec:connectiontoM5}
We have proved that all supersymmetric Mink$_4\times S^2$ solutions with non equal spinor norm have an uncharged $\text{U}(1)\times \text{U}(1)$ isometry when they are in type IIB and can be lifted to M-theory when they are in IIA. We shall now prove that they all descend from Case B in M-theory - which is the M5-brane. First let us establish that Case B implies the IIA solutions.

Whenever one has a $\text{U}(1)$ isometry in M-theory one can reduce down to type IIA along it. When one reduces from eleven dimensions to ten, it is convenient to choose a frame such that
\beq
e^a_{11}= e^{-\frac{1}{3}\Phi}e^a_{10},~a=0,1,...9,~~~~ e^{10}_{11} = e^{\frac{2}{3}\Phi} (d\psi+ C_1)
\eeq
where $e^{a}_d$ is the vielbein in $d$ space-time dimensions, $\partial_{\psi}$ is the isometry direction, $\Phi$ the dilaton  and $C_1$ the potential of the RR 2-from in IIA. When one reduces, the gamma-matrix corresponding to $e^{10}_{11}$ then becomes the chirality matrix in IIA.  The non equal norm  condition in IIA, in the conventions of section \ref{sec:noneqnormIIA},  is equivalent to $\chi^{\dag}\hat\gamma^{(6)}\chi \neq 0$ and, up to an overall warping not relevant for our discussion, the six-dimensional non-chiral spinor in IIA and seven-dimensional spinor in M-theory are the same. The non equal norm condition in M-theory therefore reads $\chi^{\dag}\gamma_m^{(7)}\chi (e^{10}_{11})^m = K_m (e^{10}_{11})^m \neq 0$.
Therefore we can establish whether or not we reduce to a non-equal-norm class in IIA by studying the M-theory one-form $K$ defined in \eqref{eq:K}. Specifically, we  generate non equal norm in IIA whenever $K$ has a leg in the isometry direction we reduce on. 

There are two ways to get a round $S^2$ factor in IIA via dimensional reduction of a parent solution in M-theory. 1) The parent solution has a round $S^2$ and a \text{U}(1) isometry orthogonal to this to reduce on. 2) The parent has a \text{U}(1) fibered over $S^2$, ie an $S^3$ that may be squashed along the Hopf fiber direction only. We can immediately rules out option 2) as in this case $K\sim d\psi+ \eta+...$,  with $d\eta = \text{Vol}(S^2)$, which can never solve the one-form condition  $d(e^{2\Delta}K)=0$ - as such IIA non equal norm must descend from Case A or B in M-theory. Here the only place\footnote{The 1-form $K$ generically has legs along, $w_1$, $dy_3$ and $v$. But $dy_3$ contains no \text{U}(1) and  the supersymmetry equation for $v$ in both M-theory classes is incompatible with it having a leg along a(n uncharged) \text{U}(1) isometry direction.} the \text{U}(1) can  lie that will give rise to non equal norm in IIA is $w_1$.  As $w_1$ is only turned on when $\sin\alpha\neq 0$, we can  realise all non equal norm solutions in type IIA through reduction  from Case B in section \ref{sec: CaseB}, which is simply the M5-brane class with SO(3) rotational symmetry in it's co-dimensions, which is rather remarkable. The local form of this solution is given in \eqref{eq:M5metflux}, where  giving $w_1$ a leg along the reduction isometry is equivalent to making the local coordinate $\partial_{x_1}$ an isometry without loss of generality\footnote{This is not the only way to form a \text{U}(1) to reduce on, in general one could take
\beq
x_1=\psi+ c_1 x,~~~~~x_2= c_2\psi+  x,~~~~~dc_i=0,
\eeq
and impose that $\partial_{\psi}$ is an isometry - however in IIA $c_i$ can be turned off with a coordinate transformation and rescaling of $g_s$. One can also form a \text{U}(1) of the metric and flux by expressing $x_1,x_2$,  in polar coordinates, however this is not an isometry of the G-structure, as \eqref{eq:K} makes clear. Indeed such an isometry is charged, so reducing on it would brake supersymmetry.}. Together with the two a prior \text{U}(1)'s in section \ref{sec: CaseB} (packaged as part of a Mink$_6$ factor there) this gives three flavour \text{U}(1) isometries, which implies that two will always remain when we reduce to IIA,  just as we found  for IIB. 
\begin{figure}
	\centering
	\begin{tikzpicture}
	\node (1B) at (3,-2) [rectangle,draw] {IIB};
	\node (1A) at (-3,-2) [rectangle,draw] {IIA};
	\node (1M) at (0,1) [rectangle,draw,very thick,fill=gray!15] {Case B};
	
	\node [black,left, xshift=-8mm] at (1A) {$\begin{matrix}T^2 & \hookrightarrow & \text{M}_4, \\
 & & \downarrow \\
 & & \text{M}_2\end{matrix}$:};
%	\node [black,below, yshift=-6mm] at (1A) {IIA};
%	\node [black,below, yshift=-6mm] at (1B) {IIB};
%	\node [black,above, yshift=6mm] at (1M) {M-theory};
	%	\node [black,above, yshift=6mm] at (3M) {M-theory};
	\path (1A) edge [bend right=25,thick,->]  node [below]  {T-duality}  (1B);
	\path (1M) edge [bend right=25,thick,->,text width=3cm]  node [left,xshift=10mm,yshift=3mm]  {dimensional \\ $\, \,$reduction} (1A);
	\end{tikzpicture}
	\caption{\small Depiction of the chains of dualities leading to  Mink$_4\times S^2$ classes in type II with non equal spinor norm. The relative complexity of solutions at each step as one moves from M-theory to IIA to IIB can increase considerably, with specific choices of \text{U}(1) the duality is performed on. However all solutions are governed by the M5 Laplacian.}
	\label{fig:relations}
\end{figure}
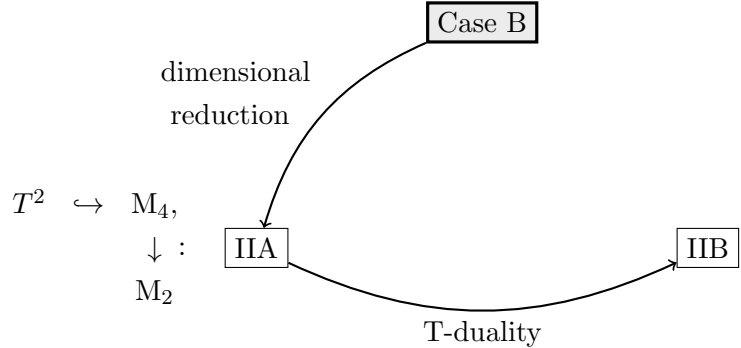

We have shown that all non equal norm solutions in IIA  descend from Case B in M-theory and all such type II solutions come with a flavour $\text{U}(1)\times \text{U}(1)$ isometry. Naively, one might then conclude that all the IIB solutions follow from IIA via T-duality - but there is another possibility to rule out. Much like option 2) for the M-theory reduction, round $S^2$ in IIB could follow by T-dualising on the Hopf fiber of a squashed $S^3$. Such a solution in IIB would necessarily have
\beq
H_1 \sim d\psi\wedge \text{Vol}(S^2),
\eeq
but fortunately for us, this was already ruled out for non equal norm in the previous section by the necessary condition $k^a(H_1)_a=0$.

We have shown that the parent or master class of all non equal norm Mink$_4\times S^2$ solutions in type II is an M5-brane with $SO(2)\times SO(3)$ rotational symmetry in it's co-dimensions. All solutions in IIA and IIB are then generated from this by a chain of dualities depicted in Fig \ref{fig:relations}. However, we should stress this does not mean that the type II classes are completely trivial - there are several  distinct ways to reduce to type IIA on a \text{U}(1) subgroup of the available $\text{U}(1)\times \text{U}(1)\times \text{U}(1)$ uncharged isometry, and yet further ways to T-dualise inside the residual $\text{U}(1)\times \text{U}(1)$ to get to IIB - the result of this chain of dualities can potentially end up being quite complicated. Rather we would like to stress that if one is interested in constructing a non equal norm Mink$_4\times S^2$ solution in type II, this is best done from the M5-brane perspective.

In the next section we shall provide master classes for the remaining Mink$_4\times S^2$ solutions in type II, namely those with equal spinor norm.

\section{Master classes for Mink$_4\times S^2$ }\label{sec:master}
In \cite{Macpherson:2016xwk,Apruzzi:2018cvq}, Mink$_4\times S^2$ solutions with equal spinor norm where classified. They fall into three distinct classes characterised by the minimum number of uncharged \text{U}(1) isometries the internal four-manifold M$_4$ contain a priori\footnote{This is in turn  related to which inner products of the two independent  non-chiral four-dimensional spinors are assumed to be non vanishing, see \cite{Macpherson:2016xwk,Apruzzi:2018cvq} for details.}. Specifically one has
\beq
\begin{matrix}\text{Case I:}~~\text{M}_4 & = & \text{M}_2\times \Sigma_2, \\
 & & ~ \\
 & & ~
\end{matrix}~~~~~~\begin{matrix}\text{Case II:}~~S^1 & \hookrightarrow & \text{M}_4, \\
 & & \downarrow \\
 & & \text{M}_3
\end{matrix}~~~~~~\begin{matrix}\text{Case III:}~~T^2 & \hookrightarrow & \text{M}_4. \\
 & & \downarrow \\
 & & \text{M}_2
\end{matrix}\nn
\eeq 
Class I has no a priori isometries on M$_4$, for  Class II one generically has a \text{U}(1) bundle over a three-manifold, while generic solutions in Class III  have a $T^2$ bundle over a two-manifold  - they can be found in sections 4.3, 4.4 and 4.5 of \cite{Macpherson:2016xwk,Apruzzi:2018cvq} for IIA and IIB respectively. Cases II and III contain one and two constant parameters respectively - when they are set to zero one reduces to classes that have local  Mink$_5\times S^2$  and Mink$_6\times S^2$ factors. More surprising is that all solutions in class III, all in IIB class II, and many in class II IIA, are governed by the same PDE's irrespective of whether or not these parameters are turned on - such generic solutions in these classes can be viewed as parametric deformations of un-fibered Mink$_{4}\times S^1\times S^2$ and Mink$_{4}\times T^2\times S^2$ solutions respectively, where the first two factors share a common warping in each instance\footnote{When this is true, locally, there is no difference between Mink$_{4}\times T^n$ and Mink$_{4+n}$. However, for the parametric deformations the warp factors are no longer the same and $T^n$ becomes fibered over a base. As such it makes more sense physically for only the Mink$_4$ directions to be non compact in these cases.}. 

The existence of parametric deformations raise an obvious question - is some sort of duality at play? Two obvious candidates are formal U-dualities of the type in \cite{Maldacena:2009mw} and T-s-T transformations \cite{Lunin:2005jy} - which are both solution generating techniques involving chains of string dualities and coordinate transformations (specifically shifts between \text{U}(1) isometries) that do not commute with these dualities. As we shall see this is enough to explain case III and case II in IIB,  but case II in IIA is more subtle.

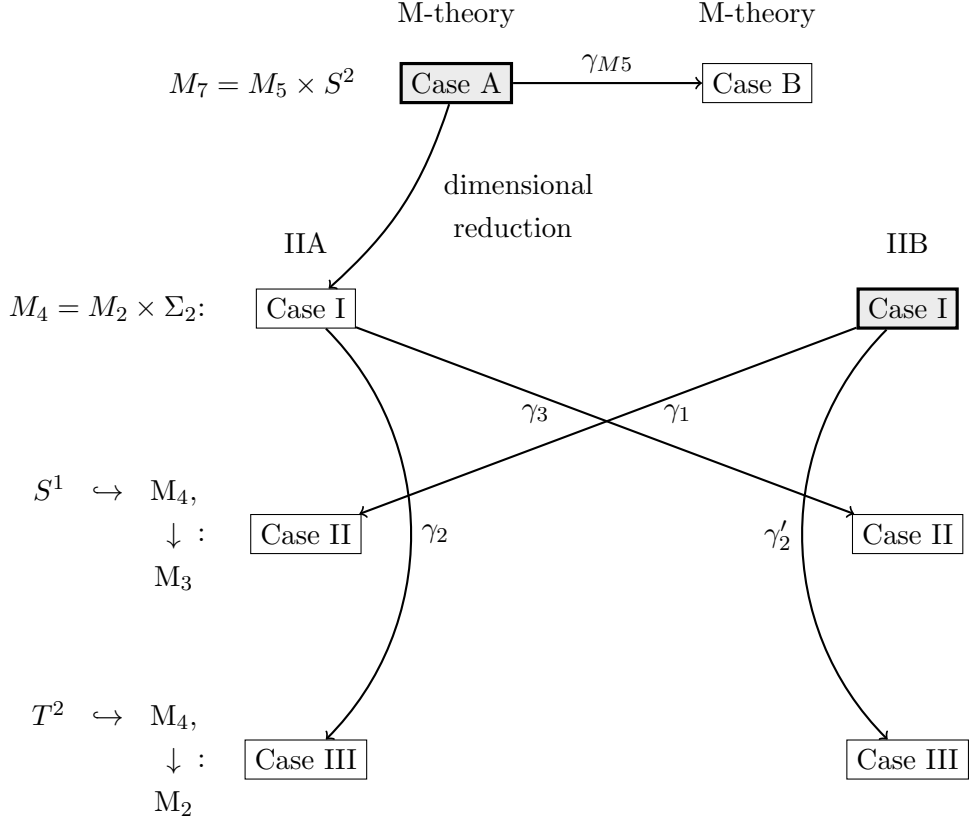
\begin{figure}
	\centering
	\begin{tikzpicture}
	\node (1B) at (4,3) [rectangle,draw,very thick,fill=gray!15] {Case I};
	\node (2B) at (4,0) [rectangle,draw] {Case II};
	\node (3B) at (4,-3)[rectangle,draw] {Case III};
	\node (1A) at (-4,3) [rectangle,draw] {Case I};
	\node (2A) at (-4,0) [rectangle,draw] {Case II};
	\node (3A) at (-4,-3)[rectangle,draw] {Case III};
	\node (1M) at (-2,6) [rectangle,draw,very thick,fill=gray!15] {Case A};
	%	\node (2M) at (5,6)[rectangle,draw] {Case C};
		\node (3M) at (2,6) [rectangle,draw] {Case B};
	
	\node [black,left, xshift=-12mm] at (1A) {$M_4=M_2 \times \Sigma_2$:};
	\node [black,left, xshift=-12mm] at (2A) {$\begin{matrix}S^1 & \hookrightarrow & \text{M}_4, \\
		& & \downarrow \\
		& & \text{M}_3\end{matrix}$:};
	\node [black,left, xshift=-12mm] at (3A) {$\begin{matrix}T^2 & \hookrightarrow & \text{M}_4, \\
		& & \downarrow \\
		& & \text{M}_2\end{matrix}$:};
	\node [black,left, xshift=-12mm] at (1M) {$M_7=M_5 \times S^2$};
	%	\node [black,right, xshift=12mm] at (2M) {$M_7=M_4 \times \widetilde{S}^3$};
	\node [black,above, yshift=6mm] at (1A) {IIA};
	\node [black,above, yshift=6mm] at (1B) {IIB};
	\node [black,above, yshift=6mm] at (1M) {M-theory};
	%	\node [black,above, yshift=6mm] at (2M) {M-theory};
		\node [black,above, yshift=6mm] at (3M) {M-theory};
	\path (1A) edge [bend left=45,thick,->]  node [right]  {$\gamma_2$}  (3A);
	\path (1A) edge [thick,->]  node [left,xshift=-6mm,yshift=1mm]  {$\gamma_3$}  (2B);
	\path (1B) edge [bend right=45,thick,->]  node [left]  {$\gamma_2'$}  (3B);
	\path (1B) edge [thick,->]  node [right,xshift=6mm,yshift=1mm]  {$\gamma_1$} (2A);
	%	\path (2A) edge [bend left=25,thick,->]  node [below,yshift=-4mm,xshift=3mm]  {$F_0=0$ lift} (2M);
	%	\path (3A)  edge [bend right=25,thick,->]  node [above,yshift=6mm,xshift=18mm]  {$F_0=0$ lift} (2M);
	\path (1M) edge [bend left=15,thick,->,text width=3cm]  node [right,xshift=4mm]  {dimensional \\ $\, \,$reduction} (1A);
	\path (1M) edge [thick,->]  node [above]  {$\gamma_{M5}$} (3M);
	%	\path (1A) edge [bend right=15,thick,->]  node [left]  {$\gamma_3$} (3M);
	\end{tikzpicture}
	\caption{Depiction of the chains of dualities leading to the various Mink$_4\times S^2$ class in type II with equal spinor norm. Cases A in M-theory and I in IIB (shaded grey) are the truly fundamental master cases from which all else can be generated. Specifically $\gamma_1$ represent a transformations where one performs a diffeomorphism mixing the $M_2$ and $\Sigma_2$ factors in IIB case I, and introducing a parameter $c_0$. One then imposes an isometry and T-dualising on it. $\gamma_2,\gamma_2'$ represent the following: impose $\Sigma_2=T^2$ in case I and T-dualise on both directions, lets call this case III$_0$, - one then performs a formal U-duality followed by a T-s-T transformation with $T^2$ - which generates a two-parameter deformation of case III$_0$ governed by the same PDE's, i.e. case III.  $\gamma_3$ represents imposing an isometry in the $\Sigma_2$ factor of IIA case I, then T-dualising to get to IIB. One then performs a formal U-duality on the spacial directions of Mink$_4$. Finally, as explained in the end of section \ref{sec: CaseB}, the physical fields (but not the spinor) of Case B can be obtained from Case A by specialising the vielbein to a M5-brane solution with some rotational symmetry in the co-dimensions; this operation in the figure is named $\gamma_{M5}$. }
	\label{fig:relations2}
\end{figure}

In the previous section we established that all solutions in type II supergravity with non equal norm can be obtained from the M5-brane, by reduction and T-duality -  so one can view the M5-brane as a master class for all such solutions. In this section, in addition to explaining the origin of the parametric deformations, we establish master classes for equal norm in type II. The first of these is M-theory Case A in section \ref{sec: CaseA}, the second is class I in IIB - note that after imposing a \text{U}(1) isometry in the M-theory class and reducing on it one end up in class I in IIA. We shall quote these solutions in the next sub-sections for convenience. All other cases can be generated from these via certain chains of dualities, as we shall show - the maps between solutions are also summarised in Fig \ref{fig:relations2}.
\subsection{Master class in type IIB: case I in IIB}
The master class in Type IIB  has an NS sector that takes the local form
\begin{align}\label{eq: conformalcalmet}
&ds^2 =e^{2A}ds^2(\text{Mink}_4) +  e^{- 2 A} \bigg(\frac{1}{f}(d x_1^2 + d x_2^2 + x^2_2 d s^2(S^2))\bigg) + ds^2(\Sigma_2),\nn\\[2mm]
&ds^2(\Sigma_2)=e^{- 2 A} (dx_3^2+dx_4^2),~~~~B=g \bigg(\text{Vol}(S^2) + \frac{d x_1 \wedge d x_2}{x_2^2}\bigg),~~~~e^{-\Phi}=f,
\end{align}
where $e^{2A}$ depends on all four-dimensional local coordinates $x_i$, and $f,g$ have support on $\Sigma_2$ only. The RR sector is generically non trivial, with $F_1,F_3,F_5$ all turned on, and can be found in  \cite{Apruzzi:2018cvq}. Solving the Bianchi identities of these fluxes (away from localised sources) requires the following PDE's to be solved
\begin{subequations}
\begin{align}
&\Box_2 f = 0 , ~~~  \Box_2 (fg) = 0,~~~ \Box_2= \partial_{x_3}^2+ \partial_{x_4}^2,\label{eq:CalabBianchi1}\\[2mm]
&\partial_{x_1}^2 (e^{-4A}) + \frac{1}{x_2^2} \partial_{x_2} ( x_2^2 \partial_{x_2} (e^{-4A}))  + \Box_2(e^{-4A} f^{-1}) +\frac{1}{x_2^4} \Box_2 (f g^2) = 0\label{eq:CalabBianchi2}.
\end{align}
\end{subequations} 
The six dimensions orthogonal to Mink$_4$ in-fact support a conformal Calabi--Yau. Clearly by imposing that one of $\partial_{x_3},\partial_{x_4}$ is an isometry of all fields and by T-dualising on it we end up with a local Mink$_5\times S^2$ class in IIA - this exhausts this class. If we make both isometries (and $g$ constant) then  T-dualise on both we exhaust the Mink$_6\times S^2$ solutions in IIB, there is just the flat space D5 brane with rotational symmetry in it's condemnations.  Of course a solution  still has a \text{U}(1) isometries when $g$ is a polynomial of order one in $(x_3,x_4)$  -  when this is the case we can gauge transform such that $B$ respects the isometry, but has a leg in it. This generically leads to the \text{U}(1) becoming fibered over the $S^2$ in the T-dual - thus one can additionally generate classes of solution with Minkowski factors and a squashed 3-sphere in this fashion.
\subsection{IIA reduction of master class in M-theory: case I in IIA }
Case I in IIA can be generated from M-theory by imposing a \text{U}(1) isometry within the SU(2)-structure of the M-theory class in  section \ref{sec: CaseA} and reducing on it. This class is rather more complicated than it's IIB counter-part, with the six-dimensional space orthogonal to Mink$_4$ supporting an orthogonal SU(2)-structure. The local form of the NS sector in this class is
\begin{align}\label{eq:IIACaseINS}
&ds^2= e^{2A}ds^2(\text{Mink}_4) + e^{-4A+2\Phi}\bigg(dx_2^2+ x_2^2 ds^2(S^2)\bigg)+  e^{4A-2\Phi}\bigg(dx_1+ B_0 e^{-4A+2\Phi}dx_2\bigg)^2+ds^2(\Sigma_2),\nn\\[2mm]
&ds^2(\Sigma_2)= e^{-2A}(dx_3^2+dx_4^2),~~~B=x_2^2 e^{-4A+2\Phi}B_0\text{Vol}(S^2),
\end{align}
where $ e^{2A}, e^{\Phi}, B_0$ each generically depends on all of the four-dimensional local coordinates $x_i$. Supersymmetry additionally requires the following PDE's to be solved
\begin{align}\label{eq: BPS ba10}
&\partial_{x_2}\left(e^{2A-2\Phi}\right)=\partial_{x_1}\left(e^{-2A}B_0\right),\nn\\[2mm]
&\partial_{x_2}\left(x_2^2e^{-2A}B_0\right)= \partial_{x_1}\left(x_2^2e^{-6A+2\Phi}(1+ B_0^2)\right).
\end{align} 
The RR sector is rather involved, but can be found in \cite{Macpherson:2016xwk}, however there is no Romans mass. Ensuring that the fluxes obey the correct Bianchi identities imposes another set of PDE's
\begin{align}\label{eq: 4dbianchis}
&\partial^2_{x_3} (e^{2A-2\Phi})+\partial^2_{x_4} (e^{2A-2\Phi})+\partial^2_{x_1}(e^{-4A})=0,\nn\\[2mm]
&\partial^2_{x_3} (e^{-2A}B_0)+\partial^2_{x_4} (e^{-2A}B_0)+\partial_{x_1} \partial_{x_2}(e^{-4A})=0,\nn\\[2mm]
&\partial^2_{x_3}(x_2^2 e^{-6A+2\Phi}(1+B_0^2))+\partial^2_{x_4}(x_2^2 e^{-6A+2\Phi}(1+B_0^2))+ \partial_{x_2}(x_2^2\partial_{x_2}(e^{-4A}))=0.
\end{align}
One can generate the Mink$_5\times S^2$ and Mink$_6\times S^2$ cases in IIB and IIA  respectively by imposing that  $\Sigma_2$ contains isometry directions, that $dB$ has no leg(s) in, and then T-dualising as before. 
\subsection{Generating case II}
In this section we shall establish how Case II in IIA and then IIB are realised from Case I in IIB and IIA respectively. The IIA case is a little subtle as generic solutions in this class are not in general related to the their Mink$_5$  sub-case by duality. In IIB things are more simple, indeed the generic solutions are essentially the U-dual of the  Mink$_5$ sub-case.

\subsubsection{Case II in IIA}
It was already observed  in \cite{Macpherson:2016xwk} that when one T-dualises on the \text{U}(1) isometry of Case II in IIA, one is mapped to a conformal Calabi--Yau class  - this is contained in case I of IIB. To see this, let us explicitly generate case II in IIA from it. We start with \eqref{eq: conformalcalmet} and perform the coordinate transformation
\beq\label{eq: diffeo1}
x_1\to x_1-  c_0 x_4
\eeq
 the NS sector of case I in IIB then takes the form 
\begin{align}\label{eq:NSshift}
d s^2 &=e^{2A}ds^2(\text{Mink}_4)+ e^{- 2 A}  \left(\frac{1}{f}\big(b^2 d x_1^2 + d x_2^2 + x^2_2 d s^2(S^2)\big) + d x_3^2 + \frac{1}{b^2}\big(d x_4+ \frac{b a_2}{\sqrt{f}} dx_1\big)^2\right),\nn\\[2mm]
B &= g {\cal C}_2-\frac{a_2 \sqrt{f}}{ b}g \frac{dx_2\wedge dx_4}{x_2^2} ,~~~ {\cal C}_2= \text{Vol}(S^2) + \frac{d x_1 \wedge d x_2}{x_2^2},
\end{align}
with the dilaton unchanged and where
\beq
b^2+a_2^2=1,~~~~ \frac{a_2\sqrt{f}}{b}=c_0.
\eeq
If we now impose that $\partial_{x_4}$ is an isometry, we have
\beq
f(x_3,x_4)\to g(x_3),\quad g(x_3,x_4)\to g(x_3),\quad A(x_1,x_2,x_3,x_4)\to A(x_1,x_2,x_3)
\eeq
and the PDE's governing the system become
\begin{subequations}
\begin{align}
&\partial_{x_3}^2 f = 0 , ~~~  \partial_{x_3}^2 (fg) = 0,\label{eq:CalabBianchi3}\\[2mm]
&\frac{1}{b^2}\partial_{x_1}^2 (e^{-4A}) + \frac{1}{x_2^2} \partial_{x_2} ( x_2^2 \partial_{x_2} (e^{-4A}))+\partial_{x_3}^2 (f^{-1}e^{-4A})  +\frac{1}{x_2^4} \partial_{x_3}^2 (f g^2) = 0\label{eq:CalabBianchi4},
\end{align}
\end{subequations}
i.e. exactly what one has for case II in IIA (cf. (5.59)- (4.60) of \cite{Macpherson:2016xwk}).  If we now T-dualise on  $\partial_{x_4}$, the Mink$_4$ part of the metric is unchanged and the rest of the NS sector becomes 
\begin{align}\label{eq:NSshiftT}
ds^2_6&=e^{2A}b^2\bigg(dx_4- \frac{a_2 g \sqrt{f}}{b x_2^2} dx_2\bigg)^2+ \frac{e^{-2A}}{f}\bigg(b^2 dx_1^2+ dx_2^2 + f dx_3^2+ x_2^2 ds^2(S^2)\bigg), \nn\\[2mm]
B&= g{\cal C}_2 + \frac{a_2 b}{\sqrt{f}} dx_1\wedge \bigg(dx_4- \frac{a_2 g \sqrt{f}}{b x_2^2}dx_2\bigg),~~~~ e^{-\Phi_{IIA}}= \frac{f}{b} e^{-A}.
\end{align}
which reproduces the case II NS sector in IIA presented in \cite{Macpherson:2016xwk} - we have checked that the same is true for the RR fluxes. After the T-duality the effect of \eqref{eq: diffeo1} can no longer be turned off with a coordinate transformation so generic \eqref{eq:NSshiftT} is a deformation of it's $c_0=0$ limit. However the PDEs with $c_0=0$  differ from the deformed case generically, due to the $b^{-2}$ factor in \eqref{eq:CalabBianchi4} that appears only in the later - which means this procedure is not generically a duality. An exception is if we assume that $\partial_{x_3}$ is also an isometry, then \eqref{eq:CalabBianchi4} is the same for all values of $c_0$. This is because when $\partial_{x_3}$ is also an isometry the diffemorphism + T-duality procedure reproduces the effect of performing a T-s-T transformation on the $c_0=0$ limit of this case. %This is not the only way to make the PDE's coincide though, indeed if $b=$ constant, we can remove it from the PDEs by rescaling $x_1$, the result is a parametric deformation, with the same PDE's for all $c_0$, but with $x_3$ no longer an isometry, so that the procedure no longer merely replicates T-s-T.

\subsubsection{Case II in IIB}
To generate case II in IIB one starts from case I in IIA. One first needs to make one direction  in $\Sigma_2$ an isometry to T-dualise on, we will take $\partial_{x_4}$ ($\partial_{x_3}$ is physically equivalent). The NS sector \eqref{eq:IIACaseINS} is then locally mapped to
\begin{align}\label{eq:IIACaseINS}
ds^2&= e^{2A}ds^2(\text{Mink}_4) + ds^2(\text{M}_6),~~~~B=x_2^2 e^{-6A+2\Phi}B_0\text{Vol}(S^2),\\[2mm]
\nn
ds^2(\text{M}_6)&=e^{2A}ds^2(S^1)+e^{-6A+2\Phi}\bigg(dx_2^2+ x_2^2 ds^2(S^2)\bigg)+  e^{6A-2\Phi}\bigg(dx_1+ B_0 e^{-6A+2\Phi}dx_2\bigg)^2+e^{-2A}dx_3^2,\nn
\end{align}
where $ds^2(S^1)=dx_4^2$, $\Phi$ is now the IIB dilaton and all functions have support in $(x_1,x_2,x_3)$ only. The  supersymmetry PDE's then become
\begin{align}\label{eq: BPS ba10asd}
&\partial_{x_2}\left(e^{4A-2\Phi}\right)=\partial_{x_1}\left(e^{-2A}B_0\right),\nn\\[2mm]
&\partial_{x_2}\left(x_2^2e^{-2A}B_0\right)= \partial_{x_1}\left(x_2^2e^{-8A+2\Phi}(1+ B_0^2)\right).
\end{align}
The RR sector also gets mapped, we will not concern ourselves with the details (see  (4.20) of \cite{Apruzzi:2018cvq} for explicit expressions), other than the fact that only $(F_1,F_3)$ are non trivial. The Bianchi identities of these fluxes require
\begin{align}\label{eq: 4dbianchisasd}
&\partial^2_{x_3} (e^{4A-2\Phi})+\partial^2_{x_1}(e^{-4A})=0,\nn\\[2mm]
&\partial^2_{x_3} (e^{-2A}B_0)+\partial_{x_1} \partial_{x_2}(e^{-4A})=0,\nn\\[2mm]
&\partial^2_{x_3}(x_2^2 e^{-8A+2\Phi}(1+B_0^2))+ \partial_{x_2}(x_2^2\partial_{x_2}(e^{-4A}))=0.
\end{align}
This is locally the Mink$_5$ limit of case II with no parametric deformation. One may then generate generic solutions in this class via a formal U-duality that introduces a parameter $c_0$ as we will now demonstrate.

One begins by T-dualising on the three space-like U(1) isometries contained in Mink$_4$ - which maps to a solution in IIA with only $(F_4,H)$ non trivial. One lifts this solution to M-theory and performs a formal boost along the M-theory circle with isometry $\partial_{z}$ 
\beq
dt \to b_1 dt+ c_0 dz,~~~~~~ dz\to b_2 dt+ dz.
\eeq
Here $c_0,b_1,b_2$ are all constants, $c_0$ is the deformation parameter, while $b_1,b_2$ should be non zero and satisfy $-b_1+ b_2 c_0>0$, but are otherwise arbitrary.  After the boost, the M-theory circle becomes fibred over $dt$ and additional components are turned on in the 4-form flux. In addition, the warp factors of the, now fibred, M-theory circle and time direction are modified. As such, when one reduces to IIA the dilaton and warp factors of the metric are modified, and one now has non trivial $(F_2,F_4,H)$ fluxes. Finally one performs an additional three T-dualities within the $\mathbb{R}^3$ factor of the solution to get back to IIB. After rescaling $dt\to -(b_1-b_2c_0) dt$ all dependence on $b_1,b_2$ drops out of the solution and one finds the new metric and dilaton
\beq\label{eq:metsd}
d\hat{s}^2= \frac{e^{2A}}{\kappa_{\perp}}ds^2(\text{Mink}_4)+ \kappa_{\perp}ds^2(\text{M}_6),~~~~e^{\hat\Phi}=e^{\Phi},\nn\\[2mm]
\eeq
where the deformed physical fields are given hats, and un-hatted fields are the original ones in IIB before the U-duality. We also introduce $(\kappa_{\|},\kappa_{\perp})$ such that
\beq
\kappa_{\|}= c_0 e^{4A-2\Phi},~~~~\kappa_{\|}^2+ \kappa_{\perp}^2=1,
\eeq
to make contact with the notation of \cite{Apruzzi:2018cvq}. The deformed fluxes are given in terms of the original ones and $(\kappa_{\|},\kappa_{\perp})$ as
\begin{align}\label{eq:fluxjf}
\hat{F}_1&=F_1,\nn\\[2mm]
\hat{F}_3&=F_3-\kappa_{\|} e^{-\Phi}\star_6 H_3,\nn\\[2mm]
\hat{H}&=dB+ \kappa_{\|} e^{\Phi}\star_6 F_3,\nn\\[2mm]  
\hat{F}_5&= \bigg(1+ \hat\star\bigg)\text{Vol}(\text{Mink}_4)\wedge \left(\frac{e^{4A-\Phi}\kappa_{\|}}{\kappa_{\perp}^2}\right),
\end{align}
where $\star_6$ is taken on the unwarped M$_6$ and $\hat\star$ on the full ten-dimensional space. It is then not hard to check that if we define $e^{2\hat A},\hat B_0$ such that  
\beq
e^{2A}=\kappa_{\perp} e^{2\hat{A}},~~~~ B_0=\kappa_{\perp} \hat B_0
\eeq
and substitute un-hatted functions for hatted ones in \eqref{eq: BPS ba10asd}, \eqref{eq: 4dbianchisasd}, \eqref{eq:metsd} and \eqref{eq:fluxjf} one precisely reproduces the metric, dilaton, fluxes and PDE's of section 4.5 in of \cite{Apruzzi:2018cvq}. Thus we have shown that case II in IIB follows from case I in IIA by first T-dualising to IIB, then performing a U-duality.
\subsection{Generating case III}
In this section we show how Case III in type IIA and IIB can be generated from one of the master systems. As this requires performing both a T-s-T transformation and U-duality, a detailed description becomes rather protracted, so we will only sketch the procedure.
 
\subsubsection{Case III in IIB}
Case III in IIB can be generated from case I in IIB by first imposing that $\Sigma_2= T^2$ so that one has two isometries to work with  and setting
\beq
g=0.
\eeq
Notice that this makes the dilaton constant  and reduces \eqref{eq:CalabBianchi2} to a Laplacian in four dimensions; indeed, if we now T-dualise on both directions in $\Sigma_2=T^2$ we are mapped to D5-brane in flat space. We can generate a double parametric deformation of the D5-brane system as follows
\begin{enumerate}
\item Formal U-duality on spatial Mink$_4$ directions\footnote{A T-duality on each of the 3 spacial directions, lift to M-theory, boost along the M-theory circle, reduce back to IIA, redo the spatial T-dualities to return to IIB.}.
\item T-dualise on $\partial_{x_4}$ and add an exact to the NS 2-form.
\item  Shift $x_3\to x_3+  \gamma x_4$.
\item T-dualise on $\partial_{x_4}$. 
\end{enumerate}
One needs to supplement this by rescaling the dilaton, Minkowski and local coordinates, but after doing this carefully one is mapped to Case III in IIB.

\subsubsection{Case III in IIA}
Case III in IIA can be generated in a similar fashion. We again fix $\Sigma_2= T^2$ and T-dualise on both \text{U}(1)'s therein. We are mapped to a Mink$_6$ system studied in \cite{Imamura:2001cr} (see also section 4.1 of \cite{Macpherson:2016xwk}), with non trivial NS and RR 0- and 2-form fluxes. We have checked that the following chain of dualities, boosts and shifts maps this to section Case III in IIA
\begin{enumerate}
\item  Four T-dualities performed on the spatial Mink$_4$ directions and one direction in $T^2$, say $x_4$.
\item  Lift to M-theory, boost along the M-theory isometry, reduce back to IIA.
\item  Shift $x_3\to x_3+ \gamma x_4$.
\item  Four more T-dualities once more on the spatial Mink$_4$ directions and $x_4$.
\end{enumerate}
Along the way some additional rescaling of coordinates is required, but this gives the general idea - i.e. Case III in IIA is generated by a combination of U-duality and T-s-T transformations.

\section{Examples}\label{sec:ex}
In this section we give some simple examples that follow from case A  in M-theory,  we exclude case B because it is merely the M5 brane - of course one AdS solution can be embedded here,  AdS$_7\times S^4/\mathbb{Z}_k$, but as this just requires solving a simple Laplace equation, we omit the details. Another result that is easy to establish is that there exist no Mink$_5\times S^2$ solution in M-theory that are not locally Mink$_6\times S^2$.  From which it follows that there also exist no AdS$_6$ solutions in M-theory.

Here we shall focus on the following: In section \ref{sec:AdS5} we show how $\mathcal{N}=2$ AdS$_5$ solutions are embedded in M-theory while in section \ref{sec:towardsMink} we provide some evidence that the squashed $S^3$ ansatz of section \ref{subsec: CaseA2} likely contains compact Mink$_4$ solutions in M-theory with non trivial fluxes.

\subsection{$\mathcal{N}=2$ AdS$_5\times S^2$ in M-theory}\label{sec:AdS5}

AdS$_5$ solutions preserving $\mathcal{N}=2$ in  M-theory were classified in \cite{Lin:2004nb} (which was shown to be an exhaustive class in \cite{OColgain:2010ev}), in this section we will show how they are embedded within M-theory class A.

The metric of these solutions is of the form
\begin{align}
ds^2&=e^{2\lambda}\bigg(4 ds^2(AdS_5) + y^2 e^{-6\lambda} ds^2(S^2)\bigg)+ \frac{4}{(1-y \partial_yD)}e^{2\lambda}\left(d\psi + V\right)^2,\nn\\[2mm]
&+ \frac{-\partial_y D}{y}e^{2\lambda}\bigg(dy^2+ e^{D} (d\hat x_1^2+ d\hat x_2^2)\bigg),\\[2mm] 
e^{-6\lambda}&=\frac{- \partial_yD}{y(1- y \partial_yD)},~~~V= \frac{1}{2}(\partial_{\hat \hat x_2} d \hat x_1- \partial_{\hat x_1}d \hat x_2),
\end{align}
where $D$ is a function of $y,x_1,x_2$ and $\partial_\psi$ is an isometry; this one together with the $S^2$ factor realises the SU$(2)\times \text{U}(1)$ R-symmetry of the $\mathcal{N}=2$ super-conformal algebra in four dimensions. Comparing this with \eqref{eq:ClassAsol} we see that if one realise AdS as
\beq
ds^2(\text{AdS}_5) = e^{2r} ds^2(\text{Mink}_4)+ dr^2,
\eeq
then one needs to fix
\begin{equation}
\rho = e^{2r} y , \qquad e^{2 \Delta} = e^{2\lambda+2r},\qquad  v= -e^{-2\lambda}(2dr+ dy).
\end{equation}
In order to get a metric that actually respects the isometries of AdS, we need the radial component to point along one of the vielbein that make up the SU(2) in such a way that the metric has no  $dr$ cross terms, and has a common warp factor for all the putative AdS directions - the SU(2) structure should also be  charged under $\partial_\psi$. After some work, one is able to establish that the following four-dimensional vielbein satisfies all the SU(2)-structure conditions
\begin{equation}
u = \sqrt{\frac{- \partial_y D}{y}} e^{\lambda + \frac{1}{2} D} (d x_1 + i d x_2), \qquad w = 2 e^{i \psi} e^{-2 \lambda} \sqrt{\frac{y}{- \partial_y D}} \left( d r + \frac{1}{2} \partial_y D d y + i (d \psi + V) \right),
\end{equation}
and yields a metric diagonal in $r$ provided that $D$ satisfies the Toda equation
\begin{equation}
(\partial_{x_1}^2+\partial_{x_2}^2 )D + \partial_y^2 e^{D} = 0,
\end{equation}
away from localised sources.
The flux $F_2$ that follows from this is then 
\begin{equation}
F_2 = e^{4 \lambda}y^{-2} \left( 2(d \psi + V)\wedge d(y^3 e^{-6 \lambda})+2y(1-y^2 e^{-6 \lambda})d V- \partial_y e^D d x_1 \wedge d x_2 \right),
\end{equation}
which reproduces the correct M-theory 4-form through  $G= e^{2C}\text{Vol}(S^2)\wedge F_2$ and obeys the Bianchi identity.

\subsection{Towards compact Mink$_4$ in M-theory with fluxes}\label{sec:towardsMink}

The ansatz of section \ref{subsec: CaseA2}, contains a squashed $S^3$ in addition to the Mink$_4\times S^2$ factor generic to all solutions we considered. As such it is a prime candidate for solutions with a Taub-NUT or Atiyah--Hitchin  factor. Metrics with Atiyah--Hitchin singularities (which are the lift of O6 planes) are a possible way to achieve compact Minkowski solutions in M-theory with non trivial fluxes. Here we show that it is indeed possible to find metrics with such singularities inside section \ref{subsec: CaseA2} with a simple example.  

The solutions in section \ref{subsec: CaseA2} are defined in terms of a single undetermined function $e^{-2K}$ that is governed by a relatively simple PDE in two variables $(\rho,y)$, namely \eqref{eq:squshedS3PDE}. The simplest way to make progress with this PDE is by a separation of variables ansatz
\beq
e^{-2K}= p(\rho)q(y).
\eeq
The PDE then reduces to two ODE's of the form
\beq
\frac{1}{\rho^2}\partial_{\rho}(\rho^2\partial_{\rho}p)+c_1 p^2=0,\qquad \partial_{y}^2 (q^2)=\frac{c_1 q}{c+y},
\eeq
where $c,c_1$ are constants. These ODE's are still difficult to solve, so in principle one can proceed numerically, however our aim is just to show the plausibility of finding compact solutions with non trivial fluxes in the squashed $S^3$ ansatz, so let us just make the most brutal non trivial assumption we can, namely
\beq
c_1=0,\qquad p=1.
\eeq
As a consequence of this
\beq
q = L\sqrt{y- b}
\eeq
for $L,b$ constants.  We necessarily have $G=0$ while the part of the metric spanned by $(\rho,S^2)$ becomes simply $\mathbb{R}^3$ or equivalently $T^3$. The metric for this solution is given (for $L=1$) by
\beq
ds^2= ds^2(\text{Mink}_4)+ ds^2(T^3)+  \frac{dy^2}{(y+c)\sqrt{y+b}}+\frac{\sqrt{y+b}}{4}ds^2(\tilde{S}^2)+\frac{y+c}{4\sqrt{y+b}}(d\psi+ \eta)^2,
\eeq
i.e. a squashed $S^3$ foliated over an interval - which is in fact bounded. One can check that
close to $y=-b$ the metric exhibits an Atiyah--Hitchin singularity while as $y$ approaches $-c$ the warp factor of $\tilde{S}^2$ becomes constant and the remaining directions vanish regularly as $\mathbb{R}^2$ in polar coordinates - thus if we assume $-c<-b$ the interval is bounded between $-c< y< -b$ and the manifold is compact.

This example of course has no fluxes turned on, but it seems likely that, even within the separation of variables ansatz, it will be possible to find similar solutions, with  $G \neq 0$, and where  $T^3\to (\rho,\text{Vol}(S^2))$ with the metric warped in terms of $\rho$. It would be interesting to pursue this - but such a detailed study is outside of the scope of this work.

\subsection*{Acknowledgements}

We would like to thank Eirik Eik Svanes and Alessandro Tomasiello for useful discussions. Additionally N. Macpherson would like to thank the IFT in Madrid,  for their warm hospitality while this work was being completed . A. Legramandi is funded in part by INFN, N. Macpherson ~is funded by the Italian Ministry of Education, Universities and Research under the Prin project ``Non Perturbative Aspects of Gauge Theories and Strings'' (2015MP2CX4) and INFN. 

\appendix

\section{Killing spinors in M-theory}\label{sec: noLukas}

In this appendix we derive some of the conditions that will be useful in section \ref{sec:Mtheoryclassification} and we prove that the M-theory case with a G$_2$ structure is actually a subcase of M-theory classification.

The eleven-dimensional supersymmetry condition can be written as
\begin{equation}
\label{11Dsusy}
D_M \eps + \frac{1}{24} (3 G \gamma_M - \gamma_M G) \eps = 0 .
\end{equation}
where $G$ is defined in \eqref{fluxdef}. The first step is to perform a decomposition from $11$ to $4+7$ dimensions
\begin{equation}
\eps =  \zeta_+ \otimes \chi +  \zeta_- \otimes \chi^c
\end{equation}
where $\zeta_+$ is a four-dimensional spinor with positive chirality,  $\zeta_+^c = \zeta_-$, and $\chi$ is a seven-dimensional one.

Evaluating \eqref{11Dsusy} on Mink$_4$ directions we get the following algebraic condition:
\begin{equation}
\label{extcondition}
\left( e^{-\Delta} \partial_{a} e^\Delta \g^a  - \frac{1}{3} i f + \frac{1}{6} F \right) \chi = 0
\end{equation}
while taking the internal index
\begin{equation}
\label{intcondition}
\left (D_{a} + \frac{1}{24} (3 F \gamma_a - \gamma_a F) - \frac{i}{12}  f \gamma_a \right) \chi = 0 .
\end{equation}
where we have chosen $\gamma_5 = i \text{Vol}_4$.
Multiplying $\eqref{extcondition}$ by $\gamma_a / 4$ we can further simplify \eqref{intcondition}:
\begin{equation}
\left (D_{a} + \frac{1}{8} F \gamma_a- \frac{i}{6}  f \gamma_a + \frac{1}{4}  e^{-\Delta} \partial^{b} e^\Delta \g_a \g_b  \right) \chi = 0 .
\end{equation}

From these conditions we can easily prove that $f = 0$, indeed contracting \eqref{extcondition}  by $\chi^\dagger$ and make the difference with the conjugate of the same expression we get
\begin{equation}
\begin{split}
&\chi^\dag \left( e^{-\Delta} \partial_{a} e^\Delta \g^a - \frac{1}{3} i f + \frac{1}{6} F \right) \chi - \overline{\chi} \left( - e^{-\Delta} \partial_{a} e^\Delta \g^a  + \frac{1}{3} i f + \frac{1}{6} F \right) \chi^c  \\
&=2i \left(e^{- \Delta}\partial_a e^{\Delta}\text{Im}( \chi^{\dag}  \gamma^a \chi)- \frac{1}{6} \text{Im}( \chi^{\dag} F \chi)- \frac{1}{3} f \right)\\
&= - \frac{2 i}{3} f \chi^\dag \chi = 0
\end{split}
\end{equation}
since the bilinears of $\chi \chi^\dag $ are real if they have degree $0,1,4,5$ and purely imaginary otherwise.

When $\chi^c = \chi$, we can only define a $G_2$ structure on M$_7$.  From the sum and the difference of \eqref{extcondition} with its conjugate we have that $F\chi=0$ and $\partial_a \Delta \gamma^a \chi = 0$. From the second condition we get:
\begin{equation}
(\partial_a \Delta \gamma^a)^2 \chi = (\partial_a \Delta \partial^a \Delta) \chi = 0
\end{equation}
which implies $ \partial^a \Delta  = 0$ since a sum of squares vanishes if an only if each individual term in the sum does. Moreover from \eqref{intcondition}
\begin{equation}
D_a \chi + D_a \chi^c = 2 D_a \chi = 0
\end{equation}
and then we can set $F = 0$. So the internal seven dimensional-manifold actually has G$_2$ holonomy and no warping factor on Mink$_4$. Although such solutions may exist with M$_7$=$S^2\times$M$_5$ we need not consider them explicitly; indeed there are no Killing spinors on $S^2$ such that $\xi= \xi^c$, and there is no invariant form on $S^2$ that maps the Killing spinor to it's Majorana conjugate - as such  imposing  $\chi^c = \chi$  makes $\chi$ contain two separable five-dimensional systems, one coupling to $\xi$ and one to $\xi^c$ - this just imposes additional constraints on the five-dimensional system that follows from $\chi^c \neq \chi$, so cases with G$_2$ holonomy are special cases of the systems we consider in the main text. We thus restrict ourselves to SU(3) structure case $\chi \neq \chi^c $. 

Now let's analyse the zero-form constraints on $M_7$:
\begin{equation}
\begin{split}
\partial_a(\chi^\dag \chi) =& D_a\chi^\dag \chi +\chi^\dag D_a \chi = \frac{1}{12} \left(\chi^\dag \gamma_a F \chi  + \chi^\dag F \gamma_a  \chi   \right) \\
=& \frac{1}{2} \partial_b \Delta \left( \chi^\dag \gamma_a \gamma_b \chi  + \chi^\dag \gamma_b \gamma_a  \chi  \right) = \partial_a \Delta \chi^\dag  \chi \, ,
\end{split}
\end{equation}
from which we have
\begin{equation}
d (e^{-\Delta} \chi^\dag \chi) = 0 
\end{equation}
and so we can set, without loss of generality, $||\chi||^2 = e^{\Delta} $.
Moreover, we can also calculate
\begin{equation}
\partial_a(\overline{\chi} \chi) = - \frac{1}{6} \left(\overline{\chi} \gamma_a F \chi  - \overline{\chi} F \gamma_a  \chi   \right) = - 2 \partial_a \Delta \chi^\dag  \chi
\end{equation}
which implies
\begin{equation}
\label{7Dspinor_norms}
d(e^{2 \Delta} \overline{\chi} \chi) = 0 .
\end{equation}
This equation tells us that the phase of $ \overline{\chi} \chi$ is constant and then we can set it to zero choosing the following parametrisation:
\begin{equation}
\overline{\chi} \chi = c e^{- 2 \Delta} 
\end{equation}
where $c$ is a real constant.

Let's analyse the two zero-form equations using our ansatz \eqref{eq:7dspinor}.
The first condition in terms of the  five- and two-dimensional spinors reads:
\begin{equation}
\frac{1}{2}(\eta_1^\dagger \eta_1 + \eta_2^\dagger \eta_2  ) + \frac{y_3}{2}(\eta_1^\dagger \eta_2 + \eta_2^\dagger \eta_1  ) = e^\Delta ,
\end{equation}
to preserve the SU$(2)$ R-symmetry the warp factor cannot depend on the coordinate on the sphere, so me have
\begin{equation}
|| \eta_1||^2 + ||\eta_2||^2  = 2 e^\Delta , \qquad \text{Re}(\eta_1^\dagger \eta_2 ) = 0.
\end{equation}
On the other hand \eqref{7Dspinor_norms} reads
\begin{equation}
- (y_1 - i y_2) \overline{\eta_1} \eta_2 = c e^{-2\Delta}
\end{equation}
and then we must have 
\begin{equation}
\overline{\eta_1} \eta_2 = 0, \qquad c = 0 .
\end{equation}
These considerably simplify the supersymmetry conditions, and in particular we can restrict to \eqref{eqconditions}, as we have done in the main text. 

%It will be useful also do the reduction from $7$ to $5 + 2$ using $\chi_+ = \chi$ as in \eqref{spinordecomposition}. The first equation in terms of the $5$-dimensional spinor reads:
%\begin{equation}
%\begin{split}
%&e^{-\Delta} \partial_{m} e^\Delta \g^m \eta_2 + \frac{i}{6} F_2 \eta_2 + \frac{1}{6} F_4 \eta_1 = 0 , \\
%&e^{-\Delta} \partial_{m} e^\Delta \g^m \eta_1 + \frac{i}{6} F_2 \eta_1 + \frac{1}{6} F_4 \eta_2 = 0 . 
%\end{split}
%\end{equation}
%The second equation splits in two pair of equations depending on whether $a$ lies along $S^2$ ($a = i$) or $M_5$ ($a = m$).
%In the first case we get:
%\begin{equation}
%\begin{split}
%& e^{-C} \partial_{m} e^C \gamma^m \eta_2 - \frac{i}{3} F_2 \eta_2 + \frac{1}{6} F_4 \eta_1 + i e^C \eta_1 = 0 , \\
%& e^{-C} \partial_{m} e^C \gamma^m \eta_1 - \frac{i}{3} F_2 \eta_1 + \frac{1}{6} F_4 \eta_2  - i e^C \eta_2 = 0 ,
%\end{split}
%\end{equation}
%while in the second one:
%\begin{equation}
%\begin{split}
%&D_{m} \eta_1 + \frac{1}{8} (i F_2 \gamma_m \eta_1 + F_4 \gamma_m \eta_2) - \frac{1}{24} \gamma_m  (i F_2 \eta_1 + F_4 \eta_2) = 0 \\
%&D_{m} \eta_2 + \frac{1}{8} (i F_2 \gamma_m \eta_2 + F_4 \gamma_m \eta_1) - \frac{1}{24} \gamma_m  (i F_2 \eta_2 + F_4 \eta_1) = 0 
%\end{split}
%\end{equation}
\section{Mink$_4\times S^2$ Killing spinor Conditions in type IIB}\label{sec:AppIIB}
In this appendix we derive the four-dimensional Killing spinor conditions a supersymmetric solutions of type IIB on Mink$_4\times S^2\times $M$_4$ must obey.
In the conventions of \cite{Kelekci:2014ima} the type IIB supersymmetry conditions are
\begin{subequations}
\begin{align}
&\partial\Phi\epsilon_1-\frac{1}{2}H\epsilon_1+e^{\Phi}\bigg(F_1+\frac{1}{2}F_3\bigg)\epsilon_2=0,\label{eq:IIB10spinors1}\\[2mm]
&\partial\Phi\epsilon_2+\frac{1}{2}H\epsilon_2+e^{\Phi}\bigg(-F_1+\frac{1}{2}F_3\bigg)\epsilon_1=0,\label{eq:IIB10spinors2}\\[2mm]
&\bigg(\nabla_{M}-\frac{1}{4}H_M\bigg)\epsilon_1-\frac{e^{\Phi}}{8}\bigg(F_1+F_3+\frac{1}{2}F_5\bigg)\Gamma_M\epsilon_2=0,\label{eq:IIB10spinors3}\\[2mm]
&\bigg(\nabla_{M}+\frac{1}{4}H_M\bigg)\epsilon_1-\frac{e^{\Phi}}{8}\bigg(-F_1+F_3-\frac{1}{2}F_5\bigg)\Gamma_M\epsilon_1=0,\label{eq:IIB10spinors4}
\end{align}
\end{subequations}
where $\epsilon_i$ are two positive chirality Majorana spinors. If we seek a solution that contains a Mink$_4$ factor foliated over the internal space by a warp factor $e^{2A}$, we can decompose the ten-dimensional Clifford algebra as
\begin{equation}
\Gamma_\mu = e^A \gamma^{(4)}_\mu \otimes \mathbb{I}, \qquad \Gamma_a = \gamma^{(4)}_5 \otimes \gamma^{(6)}_a , \qquad \mu = 0, \dots 3, \quad a= 4,\dots, 9 
\end{equation}
and the ten-dimensional spinors as
\beq
\epsilon_i = \zeta_+\otimes \chi^i_++\text{m.c}
\eeq
where $\zeta_+,\chi^i_+$ are positive chirality spinors in four- and six- dimensions respectively, and \eqref{eq:IIB10spinors1}-\eqref{eq:IIB10spinors4} become
\begin{subequations}
\begin{align}
&\partial A\chi^1_++ \frac{e^{\Phi}}{4} \bigg(F_1+F_3+\tilde{F_5}\bigg)\chi^2_+=0,\label{eq:IIB6spinors1}\\[2mm]
&\partial A\chi^2_++\frac{e^{\Phi}}{4} \bigg(-F_1+F_3-\tilde{F_5}\bigg)\chi^1_+=0,\label{eq:IIB6spinors2}\\[2mm]
&\partial\Phi\chi^1_+-\frac{1}{2}H\chi^1_++e^{\Phi}\bigg(F_1+\frac{1}{2}F_3\bigg)\chi^2_+=0,\label{eq:IIB6spinors3}\\[2mm]
&\partial\Phi\chi^2_++\frac{1}{2}H\chi^2_++e^{\Phi}\bigg(-F_1+\frac{1}{2}F_3\bigg)\chi^1_+=0,\label{eq:IIB6spinors4}\\[2mm]
&\bigg(\nabla_{a}-\frac{1}{4}H_a\bigg)\chi^1_+-\frac{e^{\Phi}}{8}\bigg(F_1+F_3+\tilde{F_5}\bigg)\gamma^{(6)}_a\chi^2_+=0,\label{eq:IIB6spinors5}\\[2mm]
&\bigg(\nabla_{a}+\frac{1}{4}H_a\bigg)\chi^2_+-\frac{e^{\Phi}}{8}\bigg(-F_1+F_3-\tilde{F_5}\bigg)\gamma^{(6)}_a\chi^1_+=0,\label{eq:IIB6spinors6}
\end{align}
\end{subequations}
where $F_1,F_3,H_3$ are purely magnetic and $\tilde{F_5}$ is the magnetic component of $F_5$.

Using \eqref{eq:IIB6spinors5} and \eqref{eq:IIB6spinors6} together with \eqref{eq:IIB6spinors1}-\eqref{eq:IIB6spinors4} multiplied by $\gamma_a$, we get the following conditions on the zero degree bilinears:
\beq\label{eq:normIIB}
\begin{split}
	&|\chi_+^1|^2+|\chi_+^2|^2= 2e^{A},~~~~|\chi_+^1|^2-|\chi_+^2|^2= 2c_- e^{-A}, \\
	&\partial_a(e^{2A-\Phi} \chi_+^{2 \dag} \chi_+^1)-e^{2A-\Phi} \partial_a (\chi_+^{1 \dag} \chi_+^2)=i e^{3A} (\star \tilde{F}_5)_a-c_- e^A (F_{1})_a.
\end{split}
\eeq
If we now assume we have an SU(2) R-symmetry parameterised by a round $S^2$ factor in our solution we have again to factorise the Clifford algebra, and we choose the following conventions: 
\begin{align}
\gamma^{(6)}_i&=\sigma_i\otimes \mathbb{I},~~~i=1,2~~~~\gamma^{(6)}_{a+2}=\sigma_3\otimes \gamma_a,~~~a=1,...,4\nn\\[2mm]
B^{(6)}&=\sigma_2\otimes B,~~~~B=B^{-1}=-B^*=-B^T,~~~~\gamma_a^*=B^{-1}\gamma_a B,
\end{align}
where $e^C$ is the warping function for $S^2$. Moreover, we have to decompose the spinors 
\begin{align}
\chi^1_+&=\frac{1}{2}\bigg[\xi\otimes (\eta^1+\eta^2)+\sigma_3\xi\otimes \hat\gamma^{(6)}(\eta^1+\eta^2)\bigg],\nn\\[2mm]
\chi^2_+&=\frac{1}{2}\bigg[\xi\otimes\hat\gamma^{(6)} (\eta^1-\eta^2)+\sigma_3\xi\otimes(\eta^1-\eta^2)\bigg],
\end{align}
and the fluxes
\beq
F=f+e^{2C}\text{Vol}(S^2)\wedge g,~~~H=H_3+e^{2C}\text{Vol}(S^2)\wedge H_1,
\eeq
one can then show that \eqref{eq:IIB6spinors1}-\eqref{eq:IIB6spinors6} is equivalent to \eqref{eq:susyIIB4d1}-\ref{eq:susyIIB4d8} in the main text.

\section{Towards solutions for the general IIA case}\label{IIA cases}
In this appendix we will find a large ansatz that allows us to simplify Case I in IIA  in such a way to have a diagonal metric. In this case supersymmetry is implied by the following equations:
\begin{subequations}
	\label{IIA_susy}
	\begin{align}
	&d(e^A w) = 0, \quad d(e^{2A+C-\Phi}) + e^{2A-\Phi} v_2 = 0, \quad d(e^{-2A + \Phi}(v_1 + B_0 v_2)) = 0 \\ \label{IIA_susy2}
	&d(e^{-\Phi}v_1) \wedge w \wedge \overline{w} = 0, \quad d(e^{2C-\Phi}(B_0 v_1 - v_2)) \wedge w \wedge \overline{w} = 0, \quad B_2 = 0.
	\end{align}
\end{subequations}
In the present case equations \eqref{IIA_susy} automatically give us a definition of a diagonal vielbein except for the component $v_1$, which we diagonalise by hand with the introduction of an arbitrary function $\mu$:
\begin{equation}
v_1 = e^{\Phi + \mu} d x_1, \quad v_2 = - e^{-2A+\Phi} d x_2, \quad w_1 = e^{-A} d x_3, \quad w_2 = e^{-A} d x_4 .
\end{equation}
where $x_2 = e^{2A + C - \Phi}$. The internal metric thus reads
\begin{equation}
\label{eq:metIIAansatz}
d s^2_6 = e^{-4A + 2 \Phi} (d x_2^2 + x_2^2 d s^2(S^2)) + e^{2 \Phi + 2 \mu} d x_1^2 + e^{-2A}(d x_3^2 + d x_4^2).
\end{equation}

It turns out to be useful define the following functions:
\begin{equation}
f = e^{-2A+2\Phi+\mu}, \quad g = e^{2C} B_0 .
\end{equation}
The other equations of \eqref{IIA_susy} imply that $f = f(x_1,x_2)$, $g=g(x_1,x_2)$ and $\mu = \mu(x_1,x_3,x_4)$. Moreover we have the following PDEs:
\begin{subequations} \label{sIIA_BPS}
	\begin{align}
	\partial_{x_1} g&= -x_2^2 \partial_{x_2} f ,\label{sIIA_BPSa} \\  
	\partial_{x_2} g &= x_2^2e^{-\mu} \partial_{x_1} (f e^{-4A-\mu}) \label{sIIA_BPSb}.
	\end{align}
\end{subequations}

The fluxes are 
\begin{subequations}
	\begin{align}
	B_2 =& g \text{Vol}(S^2), \\
	F_2 =&  (\partial_{x_4} e^{\mu} d x_3 - \partial_{x_3} e^{\mu} d x_4 ) \wedge d x_1- f^{-1} \partial_{x_1} e^{-4A} d x_3 \wedge d x_4, \\
	F_4=& B_2 \wedge F_2 +  \text{Vol}(S^2) \wedge \big( x_2^2 f (\partial_{x_3} e^{-4A-\mu} d x_4 - \partial_{x_4} e^{-4A-\mu} d x_3 ) \wedge d x_2   \nonumber \\  
	-& g (\partial_{x_4} e^{\mu} d x_3 - \partial_{x_3} e^{\mu} d x_4 ) \wedge d x_1 + (x_2^2 \partial_{x_2} e^{-4A} + g f^{-1} \partial_{x_1} e^{-4A} ) d x_3 \wedge d x_4    \big) ,
	\end{align}
\end{subequations}
and the Bianchi identities for these fluxes impose the following PDEs:
\begin{subequations}
	\begin{align}
	& \partial_{x_2} (f^{-1} \partial_{x_1} e^{-4A}) =  0 \label{eq:rest} \\
	& (\partial_{x_3}^2 + \partial_{x_4}^2 ) e^\mu + \partial_{x_1} (f^{-1} \partial_{x_1} e^{-4A}) = 0 \label{sIIA_BI2}  \\
	& f(\partial_{x_3}^2 + \partial_{x_4}^2 )e^{-4A - \mu} + \frac{1}{x_2^2} \partial_{x_2} (x_2^2 \partial_{x_2}e^{-4A}) +  \frac{1}{x_2^2} \partial_{x_2} g \,  f^{-1}  \partial_{x_1}e^{-4A}  = 0 \ \label{sIIA_BI3} .
	\end{align}
\end{subequations}

We can notice that \eqref{eq:rest} is just a restriction on the coordinate dependence of the physical fields, while \eqref{sIIA_BPS} can be used to define $H$. If this is the case, then we can define $f$ from this consistency equation 
\begin{equation}
\frac{1}{x_2^2}\partial_{x_2}( x_2^2\partial_{x_2} f)+ \partial_{x_1}(e^{-\mu}\partial_{x_1} (f e^{-4A-\mu}))=0\label{sIIA_BI1}
\end{equation} 
which is redundant whenever g gets defined rather than it's derivatives.

At this point, solving these PDEs in full generality is hard; however, this class of solutions is large enough to allow us to impose some further ansatz and still obtaining interesting intersecting-brane systems.

The first thing that needs to be addressed is that the LHS of \eqref{sIIA_BPSb} is independent of $(x_3,x_4)$, which means the RHS should also be, which is not true a priori. We can deal with this by making an ansatz for $e^{-4A}$, we find that
\begin{subequations}
	\begin{align}
	\text{Case I}: e^{-4A}&=f^{-1}e^{\mu} T(x_2,x_3,x_4) ,\\[2mm]
	\text{Case II}: e^{-4A}&=f^{-1} H(x_1,x_2) S(x_3,x_4)^2,\\
	e^{\mu}&= S(x_3,x_4)
	\end{align}
\end{subequations}
achieve the stated aims\footnote{In principle one can add a generic function $e^{\mu_0(x_1)}$ in front of the expressions for $e^{-4A}$ and $e^{\mu}$ in the second case, but it is easy to check that this can be reabsorbed into $f$ by performing a change of coordinates. Another non-trivial generalisation of the second case is to define $e^{-4A}=f^{-1}[H(x_1,x_2) S(x_3,x_4)^2+ T(x_2,x_3,x_4)]$. This case however seems to require further assumptions which so far didn't lead to something interesting.}. We will look at these cases separately in the following subsections. 

\subsection{Case I}
Here \eqref{sIIA_BPSa}-\eqref{sIIA_BPSb} imply
\beq
g'(x_1)= f_1(x_1),~~~ f= f_0(x_1)+ \frac{f_1(x_1)}{x_2} .
\eeq
The first thing we need to do is solve \eqref{eq:rest}, which becomes
\beq
\partial_{x_2}\left[\frac{1}{(x_2 f_0+ f_1 )}\partial_{x_1}\left(\frac{e^{\mu}x_2^2 T}{x_2 f_0+ f_1}\right)\right]=0.
\eeq
The obvious way to make progress with this is to do one of the following
\beq
f_0=0,~~~f_1=0,~~~\text{or}~~~f= c_0 + \frac{c_1}{x_2}~ \text{for}~ c_i~\text{constant.}
\eeq
The first two of these turn out to be subcases of the last up to a redefinition  of $x_1$ so we without loss of generality set
\beq
T=\frac{\tilde{T}(x_2,x_3,x_4)}{x_2^2 (c_1+c_2 x_2)},~~~ e^{\mu}= \tilde{S}(x_1,x_3,x_4),~~~ f(x_2)= c_0+ \frac{c_1}{x_2}
\eeq
and then \eqref{eq:rest} PDE's becomes a Youm-like condition
\beq
\partial_{x_1} \tilde{S} \partial_{x_2} \tilde{U} =0.
\eeq

\subsubsection{ $\partial_{x_1}\tilde{S}=0$: Localised D4 and smeared D6-NS5 system}
For this case $\tilde{T}$ is not particularly useful, instead we define
\beq
T=  f U(x_2,x_3,x_4),~~~ \tilde S= S(x_3,x_4).
\eeq
We are then left with the PDE's
\beq
\Box_2 S=0,~~~ f\Box_2 U+ \frac{S}{x_2^2}\partial_{x_2}(x_2^2\partial_{x_2} U)=0
\eeq
which define the system. The physical fields are defined by the solutions to these PDEs as
\beq\label{eq: physicaldata1}
e^{2A}= \frac{1}{\sqrt{U S}},~~~e^{2\Phi+2\mu}=f \sqrt{\frac{U}{S}},~~~~e^{-4A+2\Phi}=f \sqrt{\frac{S}{U}},~~~ e^{2C}B_0=c_1x_1 .
\eeq
Finally we note that the only dependence on $x_1$ appears in the NS 2-form as $B= f_1 x_1 \text{Vol}(S^2)$, which gives rise to
\beq
H= c_1 dx_1\wedge \text{Vol}(S^2),
\eeq
so $\partial_{x_1}$ is an isometry of any solution within this class. Comparing \eqref{eq: physicaldata1} and \eqref{eq:metIIAansatz} with (\cite{Youm:1999zs}, sections 4.2 and 4.5) we see that we have an ansatz for localised D4 stretched between D6-branes and ending on NS5 branes that are both smeared along $x_1$.

\subsubsection{$\partial_{x_2}\tilde{T}=0$: Smeared D4-D6-NS5 system}
Here we define
\beq
\tilde T= U(x_3,x_4),~~~ \tilde {S}= S(x_1,x_3,x_4)
\eeq
and then the remaining PDE's become
\beq
\Box_2 U =0,~~~ \Box_2 S+ U \partial_{x_1}^2S=0,
\eeq
The physical fields are defined as:
\begin{equation}
e^{2A} = \frac{1}{\sqrt{f S U}}, \quad e^{2\Phi + 2 \mu} =\sqrt{\frac{f S}{U}}, \quad e^{-4A + 2\Phi} =f^{-3/2} \sqrt{\frac{S}{U}} , \quad e^{2C}B_0 = c_1x_1
\end{equation}
We thus have a system of D4-D6-NS5 branes where the NS5 is smeared along $x_1$ while the D4 along $x_2$.

\subsection{Case II}
Here \eqref{sIIA_BPSb} becomes
\beq
\partial_{x_2} g = x_2^2e^{-\mu_0} \partial_{x_1} H
\eeq
The discussion turns out to be different if we consider the warping $A$ to depend or not from $x_1$.

\subsubsection{$\partial_{x_1}e^{-4A}=0$: Localised D4-NS5 smeared D6 system}

To make $e^{-4A}$ independent of $x_1$ we need to set 
\beq
H= f  \lambda(x_2) .
\eeq
We can solve the \eqref{sIIA_BPSa} by introducing an auxiliary function $h(x_1,x_2)$ such that
\beq
f=\partial_{x_1} h,~~~ g= -x_2^2 \partial_{x_2}h.
\eeq
We are then left with the PDE's
\beq
\partial_{x_2}(x_2^2 \partial_{x_2}\lambda)=0,~~~ \Box_2 S =0,~~~\frac{1}{x_2^2}\partial_{x_2}(x_2^2 \partial_{x_2}h)+ \lambda \partial_{x_1}^2 h =0,
\eeq
to solve, with physical fields given by
\beq\label{eq: physicaldata2}
e^{2A}= \frac{1}{S \sqrt{\lambda}},~~~e^{2\Phi+2\mu}= \frac{\partial_{x_1}h}{\sqrt{\lambda}},~~~~e^{-4A+2\Phi}=\sqrt{\lambda}\partial_{x_1} h,~~~~ e^{2C}B_0=-x_2^2\partial_{x_2} h .
\eeq
When $S=1$ this is simply the massless system of \cite{Imamura:2001cr} describing localised D6-NS5 brane intersection up to two T-dualities along $d x_3$ and $d x_4$. On the other hand when $S\neq 0$ but $\partial_{x_1} h=1$  we have a D4-D6 system where the D6 brane is smeared along $x_1$, indeed its harmonic function is $S=S(x_3,x_4)$, while the harmonic function of the D4 factorise in $\lambda S$.
therefore we can interpret our solution as a localised D4-NS5 smeared D6 system.

\subsubsection{$\partial_{x_1}e^{-4A}\neq 0$: ``Massive'' D4-D6-NS5 system}
\label{sub:massiveIIA}

We now allow $e^{-4A}$ to depend on $x_1$. Let's absorb $f^{-1}$ into $H$.
From \eqref{eq:rest} we can define $f$ from $H$ up to an arbitrary function $G(x_1) \neq 0$ (otherwise we will come back to the previous case)
\begin{equation}
f = \frac{\partial_{x_1} H}{G} . 
\end{equation}
Moreover condition \eqref{sIIA_BI2} imposes that this function has to be at least linear in $x_1$, $G = c_1 x_1 + c_2$. We then have that supersymmetry follows from the following two PDEs:
\begin{equation}
\Box_2 S = - c_1 S^2 , \qquad \frac{1}{x_2^2} \partial_{x_2} \left( x_2^2 \partial_{x_2} H \right) + \frac{1}{2} \partial_{x_1}^2 H^2 = c_1 \frac{\partial_{x_1}H^2}{c_1 x_1 + c_2} . 
\end{equation}
The fields are defined as:
\begin{equation}
\begin{split}
&e^{2A} = \frac{1}{S \sqrt{H}}, \quad e^{2\Phi + 2 \mu} = \frac{1}{\sqrt{H}G} \partial_{x_1} H , \quad  e^{-4A + 2 \Phi} = \frac{\sqrt{H}}{G} \partial_{x_1} H  \\
&\partial_{x_1} (e^{2C}B_0) = \frac{x_2^2}{G} \partial_{x_1} \partial_{x_2} H, \qquad \partial_{x_2} (e^{2C}B_0) = \frac{x_2^2}{2} \partial_{x_1} \left( \frac{\partial_{x_1} H^2}{G} \right).
\end{split}
\end{equation}
We can see that for $c_1 = 0$ this is equivalent to the previous case but with the T-dual of the massive system of localised D6-NS5 of  \cite{Imamura:2001cr}; therefore for $c_1 =0$ we have a localised ``massive'' D4-NS5 smeared D6 system. When $c_1 \neq 0$ we have something more exotic which physical interpretation is less clear, but the fluxes suggest a system of (possibly localised) D4-D6-NS5 branes.


\begin{thebibliography}{99}

%holonomy  stuff%

%\cite{Candelas:1985en}
\bibitem{Candelas:1985en}
  P.~Candelas, G.~T.~Horowitz, A.~Strominger and E.~Witten,
  ``Vacuum Configurations for Superstrings,''
  Nucl.\ Phys.\ B {\bf 258} (1985) 46.
  doi:10.1016/0550-3213(85)90602-9
  %%CITATION = doi:10.1016/0550-3213(85)90602-9;%%
  %2683 citations counted in INSPIRE as of 12 Nov 2018



%\cite{Witten:1997sc}
\bibitem{Witten:1997sc}
  E.~Witten,
  ``Solutions of four-dimensional field theories via M theory,''
  Nucl.\ Phys.\ B {\bf 500} (1997) 3
  doi:10.1016/S0550-3213(97)00416-1
  [hep-th/9703166].
  %%CITATION = doi:10.1016/S0550-3213(97)00416-1;%%
  %934 citations counted in INSPIRE as of 04 Nov 2018
	
	%\cite{Acharya:2001gy}
\bibitem{Acharya:2001gy}
  B.~S.~Acharya and E.~Witten,
  ``Chiral fermions from manifolds of G(2) holonomy,''
  hep-th/0109152.
  %%CITATION = HEP-TH/0109152;%%
  %229 citations counted in INSPIRE as of 04 Nov 2018

%\cite{Acharya:2004qe}
\bibitem{Acharya:2004qe}
  B.~S.~Acharya and S.~Gukov,
  ``M theory and singularities of exceptional holonomy manifolds,''
  Phys.\ Rept.\  {\bf 392} (2004) 121
  doi:10.1016/j.physrep.2003.10.017
  [hep-th/0409191].
  %%CITATION = doi:10.1016/j.physrep.2003.10.017;%%
  %106 citations counted in INSPIRE as of 04 Nov 2018

	
	%\cite{Douglas:2015aga}
\bibitem{Douglas:2015aga}
  M.~R.~Douglas,
  ``Calabi–Yau metrics and string compactification,''
  Nucl.\ Phys.\ B {\bf 898} (2015) 667
  doi:10.1016/j.nuclphysb.2015.04.009
  [arXiv:1503.02899 [hep-th]].
  %%CITATION = doi:10.1016/j.nuclphysb.2015.04.009;%%
  %4 citations counted in INSPIRE as of 12 Nov 2018
	
	
	
	%\cite{Acharya:2015oea}
\bibitem{Acharya:2015oea}
  B.~S.~Acharya, K.~Bożek, M.~Crispim Romão, S.~F.~King and C.~Pongkitivanichkul,
  ``SO(10) Grand Unification in M theory on a G2 manifold,''
  Phys.\ Rev.\ D {\bf 92} (2015) no.5,  055011
  doi:10.1103/PhysRevD.92.055011
  [arXiv:1502.01727 [hep-ph]].
  %%CITATION = doi:10.1103/PhysRevD.92.055011;%%
  %9 citations counted in INSPIRE as of 04 Nov 2018


%\cite{Fiset:2018huv}
\bibitem{Fiset:2018huv}
  M.~A.~Fiset,
 ``Superconformal algebras for twisted connected sums and $G_2$ mirror symmetry,''
  arXiv:1809.06376 [hep-th].
  %%CITATION = ARXIV:1809.06376;%%
  %1 citations counted in INSPIRE as of 04 Nov 2018

	
	%\cite{Kennon:2018eqg}
\bibitem{Kennon:2018eqg}
  A.~Kennon,
  ``G$_{2}$-Manifolds and M-Theory Compactifications,''
  arXiv:1810.12659 [hep-th].
  %%CITATION = ARXIV:1810.12659;%%
  %1 citations counted in INSPIRE as of 04 Nov 2018
	
	%\cite{Andriolo:2018yrz}
\bibitem{Andriolo:2018yrz}
  S.~Andriolo, G.~Shiu, H.~Triendl, T.~Van Riet, G.~Venken and G.~Zoccarato,
  ``Compact G2 holonomy spaces from SU(3) structures,''
  arXiv:1811.00063 [hep-th].
  %%CITATION = ARXIV:1811.00063;%%
	


%\cite{delaOssa:2014lma}
\bibitem{delaOssa:2014lma}
  X.~de la Ossa, M.~Larfors and E.~E.~Svanes,
  ``Exploring SU$(3)$ structure moduli spaces with integrable $G_2$ structures,''
  Adv.\ Theor.\ Math.\ Phys.\  {\bf 19} (2015) 837
  doi:10.4310/ATMP.2015.v19.n4.a5
  [arXiv:1409.7539 [hep-th]].
  %%CITATION = doi:10.4310/ATMP.2015.v19.n4.a5;%%
  %14 citations counted in INSPIRE as of 04 Nov 2018
	
	%\cite{Fiset:2017auc}
\bibitem{Fiset:2017auc}
  M.~A.~Fiset, C.~Quigley and E.~E.~Svanes,
  ``Marginal deformations of heterotic G$_{2}$ sigma models,''
  JHEP {\bf 1802} (2018) 052
  doi:10.1007/JHEP02(2018)052
  [arXiv:1710.06865 [hep-th]].
  %%CITATION = doi:10.1007/JHEP02(2018)052;%%
  %5 citations counted in INSPIRE as of 04 Nov 2018
	
	%\cite{delaOssa:2017pqy}
\bibitem{delaOssa:2017pqy}
  X.~de la Ossa, M.~Larfors and E.~E.~Svanes,
  ``The Infinitesimal Moduli Space of Heterotic G$_{2}$ Systems,''
  Commun.\ Math.\ Phys.\  {\bf 360} (2018) no.2,  727
  doi:10.1007/s00220-017-3013-8
  [arXiv:1704.08717 [hep-th]].
  %%CITATION = doi:10.1007/s00220-017-3013-8;%%
  %7 citations counted in INSPIRE as of 30 Nov 2018
	
	%\cite{delaOssa:2017gjq}
\bibitem{delaOssa:2017gjq}
  X.~de la Ossa, M.~Larfors and E.~E.~Svanes,
  ``Restrictions of Heterotic $G_2$ Structures and Instanton Connections,''
  arXiv:1709.06974 [math.DG].
  %%CITATION = ARXIV:1709.06974;%%
  %2 citations counted in INSPIRE as of 04 Nov 2018
	
	%\cite{Gauntlett:2001ur}
\bibitem{Gauntlett:2001ur}
  J.~P.~Gauntlett, N.~Kim, D.~Martelli and D.~Waldram,
  ``Five-branes wrapped on SLAG three cycles and related geometry,''
  JHEP {\bf 0111} (2001) 018
  doi:10.1088/1126-6708/2001/11/018
  [hep-th/0110034].
  %%CITATION = doi:10.1088/1126-6708/2001/11/018;%%
  %141 citations counted in INSPIRE as of 04 Nov 2018
	
	
%\cite{Friedrich:2001yp}
\bibitem{Friedrich:2001yp}
  T.~Friedrich and S.~Ivanov,
  ``Killing spinor equations in dimension 7 and geometry of integrable G(2) manifolds,''
  J.\ Geom.\ Phys.\  {\bf 48} (2003) 1
  doi:10.1016/S0393-0440(03)00005-6
  [math/0112201 [math.DG]].
  %%CITATION = doi:10.1016/S0393-0440(03)00005-6;%%
  %71 citations counted in INSPIRE as of 04 Nov 2018
	
%\cite{Gauntlett:2002sc}
\bibitem{Gauntlett:2002sc}
  J.~P.~Gauntlett, D.~Martelli, S.~Pakis and D.~Waldram,
 ``G structures and wrapped NS5-branes,''
  Commun.\ Math.\ Phys.\  {\bf 247} (2004) 421
  doi:10.1007/s00220-004-1066-y
  [hep-th/0205050].
  %%CITATION = doi:10.1007/s00220-004-1066-y;%%
  %276 citations counted in INSPIRE as of 04 Nov 2018
	
	%\cite{Gauntlett:2003cy}
\bibitem{Gauntlett:2003cy}
  J.~P.~Gauntlett, D.~Martelli and D.~Waldram,
  ``Superstrings with intrinsic torsion,''
  Phys.\ Rev.\ D {\bf 69} (2004) 086002
  doi:10.1103/PhysRevD.69.086002
  [hep-th/0302158].
  %%CITATION = doi:10.1103/PhysRevD.69.086002;%%
  %292 citations counted in INSPIRE as of 04 Nov 2018
	


	%\cite{Gibbons:1984kp}
\bibitem{Gibbons:1984kp}
  G.~W.~Gibbons,
  ``Aspects Of Supergravity Theories,''
  Print-85-0061 (CAMBRIDGE).
  %%CITATION = PRINT-85-0061 (CAMBRIDGE);%%
  %7 citations counted in INSPIRE as of 04 Nov 2018
	
	%\cite{deWit:1986mwo}
\bibitem{deWit:1986mwo}
  B.~de Wit, D.~J.~Smit and N.~D.~Hari Dass,
  ``Residual Supersymmetry of Compactified D=10 Supergravity,''
  Nucl.\ Phys.\ B {\bf 283} (1987) 165.
  doi:10.1016/0550-3213(87)90267-7
  %%CITATION = doi:10.1016/0550-3213(87)90267-7;%%
  %233 citations counted in INSPIRE as of 04 Nov 2018
	
	%\cite{Maldacena:2000mw}
\bibitem{Maldacena:2000mw}
  J.~M.~Maldacena and C.~Nunez,
  ``Supergravity description of field theories on curved manifolds and a no go theorem,''
  Int.\ J.\ Mod.\ Phys.\ A {\bf 16} (2001) 822
  doi:10.1142/S0217751X01003935, 10.1142/S0217751X01003937
  [hep-th/0007018].
  %%CITATION = doi:10.1142/S0217751X01003935, 10.1142/S0217751X01003937;%%
  %799 citations counted in INSPIRE as of 04 Nov 2018
	
	

	
	%\cite{Becker:1996gj}
\bibitem{Becker:1996gj}
  K.~Becker and M.~Becker,
 ``M theory on eight manifolds,''
  Nucl.\ Phys.\ B {\bf 477} (1996) 155
  doi:10.1016/0550-3213(96)00367-7
  [hep-th/9605053].
  %%CITATION = doi:10.1016/0550-3213(96)00367-7;%%
  %417 citations counted in INSPIRE as of 04 Nov 2018
	
	%\cite{Dasgupta:1999ss}
\bibitem{Dasgupta:1999ss}
  K.~Dasgupta, G.~Rajesh and S.~Sethi,
  ``M theory, orientifolds and G - flux,''
  JHEP {\bf 9908} (1999) 023
  doi:10.1088/1126-6708/1999/08/023
  [hep-th/9908088].
  %%CITATION = doi:10.1088/1126-6708/1999/08/023;%%
  %662 citations counted in INSPIRE as of 04 Nov 2018
	
	%\cite{Giddings:2001yu}
\bibitem{Giddings:2001yu}
  S.~B.~Giddings, S.~Kachru and J.~Polchinski,
 ``Hierarchies from fluxes in string compactifications,''
  Phys.\ Rev.\ D {\bf 66} (2002) 106006
  doi:10.1103/PhysRevD.66.106006
  [hep-th/0105097].
  %%CITATION = doi:10.1103/PhysRevD.66.106006;%%
  %1728 citations counted in INSPIRE as of 04 Nov 2018
	

		
			%\cite{Martelli:2003ki}
\bibitem{Martelli:2003ki}
  D.~Martelli and J.~Sparks,
  ``G structures, fluxes and calibrations in M theory,''
  Phys.\ Rev.\ D {\bf 68} (2003) 085014
  doi:10.1103/PhysRevD.68.085014
  [hep-th/0306225].
  %%CITATION = doi:10.1103/PhysRevD.68.085014;%%
  %127 citations counted in INSPIRE as of 04 Nov 2018
	
	%\cite{Kaste:2003zd}
\bibitem{Kaste:2003zd}
  P.~Kaste, R.~Minasian and A.~Tomasiello,
 ``Supersymmetric M theory compactifications with fluxes on seven-manifolds and G structures,''
  JHEP {\bf 0307} (2003) 004
  doi:10.1088/1126-6708/2003/07/004
  [hep-th/0303127].
  %%CITATION = doi:10.1088/1126-6708/2003/07/004;%%
  %87 citations counted in INSPIRE as of 21 Jun 2017

	
		%\cite{DallAgata:2003txk}
\bibitem{DallAgata:2003txk}
  G.~Dall'Agata and N.~Prezas,
``N = 1 geometries for M theory and type IIA strings with fluxes,''
  Phys.\ Rev.\ D {\bf 69} (2004) 066004
  doi:10.1103/PhysRevD.69.066004
  [hep-th/0311146].
  %%CITATION = doi:10.1103/PhysRevD.69.066004;%%
  %74 citations counted in INSPIRE as of 21 Jun 2017
	
		%\cite{Lukas:2004ip}
\bibitem{Lukas:2004ip}
  A.~Lukas and P.~M.~Saffin,
  ``M theory compactification, fluxes and AdS(4),''
  Phys.\ Rev.\ D {\bf 71} (2005) 046005
  doi:10.1103/PhysRevD.71.046005
  [hep-th/0403235].
  %%CITATION = doi:10.1103/PhysRevD.71.046005;%%
  %39 citations counted in INSPIRE as of 21 Jun 2017
	
		%\cite{Gauntlett:2004zh}
\bibitem{Gauntlett:2004zh}
  J.~P.~Gauntlett, D.~Martelli, J.~Sparks and D.~Waldram,
  ``Supersymmetric AdS(5) solutions of M theory,''
  Class.\ Quant.\ Grav.\  {\bf 21} (2004) 4335
  doi:10.1088/0264-9381/21/18/005
  [hep-th/0402153].
  %%CITATION = doi:10.1088/0264-9381/21/18/005;%%
  %258 citations counted in INSPIRE as of 21 Jun 2017
	

	
	%\cite{Grana:2004bg}
\bibitem{Grana:2004bg}
  M.~Grana, R.~Minasian, M.~Petrini and A.~Tomasiello,
  ``Supersymmetric backgrounds from generalized Calabi-Yau manifolds,''
  JHEP {\bf 0408} (2004) 046
  doi:10.1088/1126-6708/2004/08/046
  [hep-th/0406137].
  %%CITATION = doi:10.1088/1126-6708/2004/08/046;%%
  %256 citations counted in INSPIRE as of 04 Nov 2018
	

	%\cite{Grana:2005sn}
\bibitem{Grana:2005sn}
  M.~Grana, R.~Minasian, M.~Petrini and A.~Tomasiello,
  ``Generalized structures of N=1 vacua,''
  JHEP {\bf 0511} (2005) 020
  doi:10.1088/1126-6708/2005/11/020
  [hep-th/0505212].
  %%CITATION = doi:10.1088/1126-6708/2005/11/020;%%
  %249 citations counted in INSPIRE as of 04 Nov 2018
	
	%\cite{Haack:2009jg}
\bibitem{Haack:2009jg}
  M.~Haack, D.~Lust, L.~Martucci and A.~Tomasiello,
 ``Domain walls from ten dimensions,''
  JHEP {\bf 0910} (2009) 089
  doi:10.1088/1126-6708/2009/10/089
  [arXiv:0905.1582 [hep-th]].
  %%CITATION = doi:10.1088/1126-6708/2009/10/089;%%
  %28 citations counted in INSPIRE as of 04 Nov 2018

	
	%\cite{Lust:2010by}
\bibitem{Lust:2010by}
  D.~Lust, P.~Patalong and D.~Tsimpis,
  ``Generalized geometry, calibrations and supersymmetry in diverse dimensions,''
  JHEP {\bf 1101} (2011) 063
  doi:10.1007/JHEP01(2011)063
  [arXiv:1010.5789 [hep-th]].
  %%CITATION = doi:10.1007/JHEP01(2011)063;%%
  %34 citations counted in INSPIRE as of 04 Nov 2018
	
		%\cite{Prins:2013koa}
\bibitem{Prins:2013koa}
  D.~Prins and D.~Tsimpis,
  ``IIB supergravity on manifolds with SU(4) structure and generalized geometry,''
  JHEP {\bf 1307} (2013) 180
  doi:10.1007/JHEP07(2013)180
  [arXiv:1306.2543 [hep-th]].
  %%CITATION = doi:10.1007/JHEP07(2013)180;%%
  %20 citations counted in INSPIRE as of 04 Nov 2018
	

	%\cite{Prins:2013wza}
\bibitem{Prins:2013wza}
  D.~Prins and D.~Tsimpis,
  ``Type IIA supergravity and M -theory on manifolds with SU(4) structure,''
  Phys.\ Rev.\ D {\bf 89} (2014) 064030
  doi:10.1103/PhysRevD.89.064030
  [arXiv:1312.1692 [hep-th]].
  %%CITATION = doi:10.1103/PhysRevD.89.064030;%%
  %24 citations counted in INSPIRE as of 04 Nov 2018
	
	%\cite{Rosa:2013lwa}
\bibitem{Rosa:2013lwa}
  D.~Rosa,
  ``Generalized geometry of two-dimensional vacua,''
  JHEP {\bf 1407} (2014) 111
  doi:10.1007/JHEP07(2014)111
  [arXiv:1310.6357 [hep-th]].
  %%CITATION = doi:10.1007/JHEP07(2014)111;%%
  %15 citations counted in INSPIRE as of 04 Nov 2018
	

	

	%\cite{Macpherson:2017mvu}
\bibitem{Macpherson:2017mvu}
  N.~T.~Macpherson, J.~Montero and D.~Prins,
  ``Mink $_3\times S^3$ solutions of type II supergravity,''
  Nucl.\ Phys.\ B {\bf 933} (2018) 185
  doi:10.1016/j.nuclphysb.2018.05.021
  [arXiv:1712.00851 [hep-th]].
  %%CITATION = doi:10.1016/j.nuclphysb.2018.05.021;%%
  %3 citations counted in INSPIRE as of 04 Nov 2018
	
	%\cite{Legramandi:2018qkr}
\bibitem{Legramandi:2018qkr}
  A.~Legramandi, L.~Martucci and A.~Tomasiello,
 ``Timelike structures of ten-dimensional supersymmetry,''
  arXiv:1810.08625 [hep-th].
  %%CITATION = ARXIV:1810.08625;%%
	

	%\cite{Candelas:2014jma}
\bibitem{Candelas:2014jma}
  P.~Candelas, A.~Constantin, C.~Damian, M.~Larfors and J.~F.~Morales,
  ``Type IIB flux vacua from G-theory I,''
  JHEP {\bf 1502} (2015) 187
  doi:10.1007/JHEP02(2015)187
  [arXiv:1411.4785 [hep-th]].
  %%CITATION = doi:10.1007/JHEP02(2015)187;%%
  %14 citations counted in INSPIRE as of 04 Nov 2018
	
	%\cite{Candelas:2014kma}
\bibitem{Candelas:2014kma}
  P.~Candelas, A.~Constantin, C.~Damian, M.~Larfors and J.~F.~Morales,
  ``Type IIB flux vacua from G-theory II,''
  JHEP {\bf 1502} (2015) 188
  doi:10.1007/JHEP02(2015)188
  [arXiv:1411.4786 [hep-th]].
  %%CITATION = doi:10.1007/JHEP02(2015)188;%%
  %14 citations counted in INSPIRE as of 04 Nov 2018
	
%\cite{Freund:1980xh}
\bibitem{Freund:1980xh}
  P.~G.~O.~Freund and M.~A.~Rubin,
  ``Dynamics of Dimensional Reduction,''
  Phys.\ Lett.\ B {\bf 97} (1980) 233
   [Phys.\ Lett.\  {\bf 97B} (1980) 233].
  doi:10.1016/0370-2693(80)90590-0
  %%CITATION = doi:10.1016/0370-2693(80)90590-0;%%
  %885 citations counted in INSPIRE as of 04 Nov 2018
	
		%\cite{Apruzzi:2013yva}
\bibitem{Apruzzi:2013yva}
  F.~Apruzzi, M.~Fazzi, D.~Rosa and A.~Tomasiello,
  ``All AdS$_7$ solutions of type II supergravity,''
  JHEP {\bf 1404} (2014) 064
  doi:10.1007/JHEP04(2014)064
  [arXiv:1309.2949 [hep-th]].
  %%CITATION = doi:10.1007/JHEP04(2014)064;%%
  %89 citations counted in INSPIRE as of 04 Nov 2018
	
	
%\cite{Passias:2012vp}
\bibitem{Passias:2012vp}
  A.~Passias,
  ``A note on supersymmetric AdS$_6$ solutions of massive type IIA supergravity,''
  JHEP {\bf 1301} (2013) 113
  doi:10.1007/JHEP01(2013)113
  [arXiv:1209.3267 [hep-th]].
  %%CITATION = doi:10.1007/JHEP01(2013)113;%%
  %39 citations counted in INSPIRE as of 04 Nov 2018
	

			%\cite{Colgain:2011hb}
\bibitem{Colgain:2011hb}
  E.~O Colgain and B.~Stefanski, Jr.,
  ``A search for AdS5 X S2 IIB supergravity solutions dual to N = 2 SCFTs,''
  JHEP {\bf 1110} (2011) 061
  doi:10.1007/JHEP10(2011)061
  [arXiv:1107.5763 [hep-th]].
  %%CITATION = doi:10.1007/JHEP10(2011)061;%%
  %19 citations counted in INSPIRE as of 04 Nov 2018
	
	%\cite{Apruzzi:2014qva}
\bibitem{Apruzzi:2014qva}
  F.~Apruzzi, M.~Fazzi, A.~Passias, D.~Rosa and A.~Tomasiello,
  ``AdS$_{6}$ solutions of type II supergravity,''
  JHEP {\bf 1411} (2014) 099
   Erratum: [JHEP {\bf 1505} (2015) 012]
  doi:10.1007/JHEP11(2014)099, 10.1007/JHEP05(2015)012
  [arXiv:1406.0852 [hep-th]].
  %%CITATION = doi:10.1007/JHEP11(2014)099, 10.1007/JHEP05(2015)012;%%
  %57 citations counted in INSPIRE as of 04 Nov 2018
	
	%\cite{DHoker:2016ujz}
\bibitem{DHoker:2016ujz} 
  E.~D'Hoker, M.~Gutperle, A.~Karch and C.~F.~Uhlemann,
  ``Warped $AdS_6\times S^2$ in Type IIB supergravity I: Local solutions,''
  JHEP {\bf 1608}, 046 (2016)
  doi:10.1007/JHEP08(2016)046
  [arXiv:1606.01254 [hep-th]].
  %%CITATION = doi:10.1007/JHEP08(2016)046;%%
  %33 citations counted in INSPIRE as of 04 Nov 2018
	
	%\cite{Lin:2004nb}
\bibitem{Lin:2004nb}
  H.~Lin, O.~Lunin and J.~M.~Maldacena,
  ``Bubbling AdS space and 1/2 BPS geometries,''
  JHEP {\bf 0410} (2004) 025
  doi:10.1088/1126-6708/2004/10/025
  [hep-th/0409174].
  %%CITATION = doi:10.1088/1126-6708/2004/10/025;%%
  %690 citations counted in INSPIRE as of 04 Nov 2018
	
	%\cite{ReidEdwards:2010qs}
\bibitem{ReidEdwards:2010qs}
  R.~A.~Reid-Edwards and B.~Stefanski, jr.,
``On Type IIA geometries dual to N = 2 SCFTs,''
  Nucl.\ Phys.\ B {\bf 849} (2011) 549
  doi:10.1016/j.nuclphysb.2011.04.002
  [arXiv:1011.0216 [hep-th]].
  %%CITATION = doi:10.1016/j.nuclphysb.2011.04.002;%%
  %32 citations counted in INSPIRE as of 29 Nov 2018
	
	%\cite{DHoker:2016ujz}
\bibitem{DHoker:2016ujz}
  E.~D'Hoker, M.~Gutperle, A.~Karch and C.~F.~Uhlemann,
 ``Warped $AdS_6\times S^2$ in Type IIB supergravity I: Local solutions,''
  JHEP {\bf 1608} (2016) 046
  doi:10.1007/JHEP08(2016)046
  [arXiv:1606.01254 [hep-th]].
  %%CITATION = doi:10.1007/JHEP08(2016)046;%%
  %33 citations counted in INSPIRE as of 04 Nov 2018

	%\cite{Macpherson:2016xwk}
\bibitem{Macpherson:2016xwk}
  N.~T.~Macpherson and A.~Tomasiello,
 ``Minimal flux Minkowski classification,''
  arXiv:1612.06885 [hep-th].
  %%CITATION = ARXIV:1612.06885;%%
  %2 citations counted in INSPIRE as of 21 Jun 2017

	
	%\cite{Apruzzi:2018cvq}
\bibitem{Apruzzi:2018cvq}
  F.~Apruzzi, J.~C.~Geipel, A.~Legramandi, N.~T.~Macpherson and M.~Zagermann,
  ``Minkowski$_4$ $\times$ $S^2$ solutions of IIB supergravity,''
  Fortsch.\ Phys.\  {\bf 66} (2018) no.3,  1800006
  doi:10.1002/prop.201800006
  [arXiv:1801.00800 [hep-th]].
  %%CITATION = doi:10.1002/prop.201800006;%%
  %4 citations counted in INSPIRE as of 04 Nov 2018
	


	%\cite{Corbino:2017tfl}
\bibitem{Corbino:2017tfl}
  D.~Corbino, E.~D'Hoker and C.~F.~Uhlemann,
 ``AdS$_{2}$ × S$^{6}$ versus AdS$_{6}$ × S$^{2}$ in Type IIB supergravity,''
  JHEP {\bf 1803} (2018) 120
  doi:10.1007/JHEP03(2018)120
  [arXiv:1712.04463 [hep-th]].
  %%CITATION = doi:10.1007/JHEP03(2018)120;%%
  %5 citations counted in INSPIRE as of 04 Nov 2018
	
		%\cite{Dibitetto:2018gbk}
\bibitem{Dibitetto:2018gbk}
  G.~Dibitetto and A.~Passias,
  ``$\textrm{AdS}_{2}\times S^7$ solutions from D0 $-$ F1 $-$ D8 intersections,''
  JHEP {\bf 1810} (2018) 190
  doi:10.1007/JHEP10(2018)190
  [arXiv:1807.00555 [hep-th]].
  %%CITATION = doi:10.1007/JHEP10(2018)190;%%
  %1 citations counted in INSPIRE as of 04 Nov 2018
	
	%\cite{Dibitetto:2018ftj}
\bibitem{Dibitetto:2018ftj}
  G.~Dibitetto, G.~Lo Monaco, A.~Passias, N.~Petri and A.~Tomasiello,
  ``AdS$_3$ solutions with exceptional supersymmetry,''
  Fortsch.\ Phys.\  {\bf 66} (2018) no.10,  1800060
  doi:10.1002/prop.201800060
  [arXiv:1807.06602 [hep-th]].
  %%CITATION = doi:10.1002/prop.201800060;%%
  %1 citations counted in INSPIRE as of 04 Nov 2018
  
  %\cite{Lust:2004ig}
  \bibitem{Lust:2004ig} 
  D.~Lust and D.~Tsimpis,
  ``Supersymmetric AdS(4) compactifications of IIA supergravity,''
  JHEP {\bf 0502}, 027 (2005)
  doi:10.1088/1126-6708/2005/02/027
  [hep-th/0412250].
  %%CITATION = doi:10.1088/1126-6708/2005/02/027;%%
  %196 citations counted in INSPIRE as of 03 Dec 2018
  
 % \cite{Koerber:2008rx}
  \bibitem{Koerber:2008rx} 
  P.~Koerber, D.~Lust and D.~Tsimpis,
  ``Type IIA AdS(4) compactifications on cosets, interpolations and domain walls,''
  JHEP {\bf 0807}, 017 (2008)
  doi:10.1088/1126-6708/2008/07/017
  [arXiv:0804.0614 [hep-th]].
  %%CITATION = doi:10.1088/1126-6708/2008/07/017;%%
  %83 citations counted in INSPIRE as of 03 Dec 2018
	
		%\cite{Apruzzi:2015zna}

%\cite{Gauntlett:2005ww}
\bibitem{Gauntlett:2005ww}
  J.~P.~Gauntlett, D.~Martelli, J.~Sparks and D.~Waldram,
  ``Supersymmetric AdS(5) solutions of type IIB supergravity,''
  Class.\ Quant.\ Grav.\  {\bf 23} (2006) 4693
  doi:10.1088/0264-9381/23/14/009
  [hep-th/0510125].
  %%CITATION = doi:10.1088/0264-9381/23/14/009;%%
  %143 citations counted in INSPIRE as of 04 Nov 2018
  
  %\cite{Passias:2018zlm}
  \bibitem{Passias:2018zlm} 
  A.~Passias, D.~Prins and A.~Tomasiello,
  ``A massive class of $\mathcal{N} = 2$ AdS$_4$ IIA solutions,''
  JHEP {\bf 1810}, 071 (2018)
  doi:10.1007/JHEP10(2018)071
  [arXiv:1805.03661 [hep-th]].
  %%CITATION = doi:10.1007/JHEP10(2018)071;%%
  %5 citations counted in INSPIRE as of 03 Dec 2018
  
	
	%\cite{Couzens:2016iot}
\bibitem{Couzens:2016iot}
  C.~Couzens,
  ``Supersymmetric AdS$_{5}$ solutions of type IIB supergravity without D3 branes,''
  JHEP {\bf 1701} (2017) 041
  doi:10.1007/JHEP01(2017)041
  [arXiv:1609.05039 [hep-th]].
  %%CITATION = doi:10.1007/JHEP01(2017)041;%%
  %3 citations counted in INSPIRE as of 04 Nov 2018
	
\bibitem{Apruzzi:2015zna}
F.~Apruzzi, M.~Fazzi, A.~Passias and A.~Tomasiello,
``Supersymmetric AdS$_{5}$ solutions of massive IIA supergravity,''
JHEP {\bf 1506} (2015) 195
doi:10.1007/JHEP06(2015)195
[arXiv:1502.06620 [hep-th]].
%%CITATION = doi:10.1007/JHEP06(2015)195;%%
%44 citations counted in INSPIRE as of 04 Nov 2018

%\cite{Cordova:2018eba}
\bibitem{Cordova:2018eba} 
C.~Cordova, G.~B.~De Luca and A.~Tomasiello,
``AdS$_8$ Solutions in Type II Supergravity,''
arXiv:1811.06987 [hep-th].
%%CITATION = ARXIV:1811.06987;%%
%1 citations counted in INSPIRE as of 03 Dec 2018

	
	%\cite{Malek:2018zcz}
\bibitem{Malek:2018zcz}
  E.~Malek, H.~Samtleben and V.~Vall Camell,
  ``Supersymmetric AdS$_{7}$ and AdS$_6$ vacua and their minimal consistent truncations from exceptional field theory,''
  Phys.\ Lett.\ B {\bf 786} (2018) 171
  doi:10.1016/j.physletb.2018.09.037
  [arXiv:1808.05597 [hep-th]].
  %%CITATION = doi:10.1016/j.physletb.2018.09.037;%%
  %7 citations counted in INSPIRE as of 04 Nov 2018
	
	%\cite{DeLuca:2018zbi}
\bibitem{DeLuca:2018zbi}
  G.~B.~De Luca, A.~Gnecchi, G.~L.~Monaco and A.~Tomasiello,
  ``Holographic duals of 6d RG flows,''
  arXiv:1810.10013 [hep-th].
  %%CITATION = ARXIV:1810.10013;%%




	

	%\cite{Apruzzi:2015zna}
\bibitem{Apruzzi:2015zna}
  F.~Apruzzi, M.~Fazzi, A.~Passias and A.~Tomasiello,
  ``Supersymmetric AdS$_{5}$ solutions of massive IIA supergravity,''
  JHEP {\bf 1506} (2015) 195
  doi:10.1007/JHEP06(2015)195
  [arXiv:1502.06620 [hep-th]].
  %%CITATION = doi:10.1007/JHEP06(2015)195;%%
  %29 citations counted in INSPIRE as of 18 Jun 2017
	
	
	%\cite{Rota:2015aoa}
\bibitem{Rota:2015aoa}
  A.~Rota and A.~Tomasiello,
  ``AdS$_{4}$ compactifications of AdS$_{7}$ solutions in type II supergravity,''
  JHEP {\bf 1507} (2015) 076
  doi:10.1007/JHEP07(2015)076
  [arXiv:1502.06622 [hep-th]].
  %%CITATION = doi:10.1007/JHEP07(2015)076;%%
  %29 citations counted in INSPIRE as of 04 Nov 2018
	
	
		%\cite{Gauntlett:2002fz}
\bibitem{Gauntlett:2002fz}
  J.~P.~Gauntlett and S.~Pakis,
  ``The Geometry of D = 11 killing spinors,''
  JHEP {\bf 0304} (2003) 039
  doi:10.1088/1126-6708/2003/04/039
  [hep-th/0212008].
  %%CITATION = doi:10.1088/1126-6708/2003/04/039;%%
  %264 citations counted in INSPIRE as of 21 Jun 2017
	
		%\cite{DeLuca:2018buk}
\bibitem{DeLuca:2018buk}
  G.~B.~De Luca, G.~L.~Monaco, N.~T.~Macpherson, A.~Tomasiello and O.~Varela,
  ``The geometry of $ \mathcal{N}=3 $ AdS$_{4}$ in massive IIA,''
  JHEP {\bf 1808} (2018) 133
  doi:10.1007/JHEP08(2018)133
  [arXiv:1805.04823 [hep-th]].
  %%CITATION = doi:10.1007/JHEP08(2018)133;%%
	
	%\cite{Candelas:1984yd}
\bibitem{Candelas:1984yd}
  P.~Candelas and D.~J.~Raine,
  ``Spontaneous Compactification and Supersymmetry in $d=11$ Supergravity,''
  Nucl.\ Phys.\ B {\bf 248} (1984) 415.
  doi:10.1016/0550-3213(84)90604-7
  %%CITATION = doi:10.1016/0550-3213(84)90604-7;%%
  %56 citations counted in INSPIRE as of 18 Oct 2018
	
	%\cite{Kelekci:2014ima}
\bibitem{Kelekci:2014ima}
  O.~Kelekci, Y.~Lozano, N.~T.~Macpherson and E.~Ó.~Colgáin,
  ``Supersymmetry and non-Abelian T-duality in type II supergravity,''
  Class.\ Quant.\ Grav.\  {\bf 32} (2015) no.3,  035014
  doi:10.1088/0264-9381/32/3/035014
  [arXiv:1409.7406 [hep-th]].
  %%CITATION = doi:10.1088/0264-9381/32/3/035014;%%
  %44 citations counted in INSPIRE as of 03 Nov 2018
	
	%\cite{Maldacena:2009mw}
\bibitem{Maldacena:2009mw}
  J.~Maldacena and D.~Martelli,
  ``The Unwarped, resolved, deformed conifold: Fivebranes and the baryonic branch of the Klebanov-Strassler theory,''
  JHEP {\bf 1001} (2010) 104
  doi:10.1007/JHEP01(2010)104
  [arXiv:0906.0591 [hep-th]].
  %%CITATION = doi:10.1007/JHEP01(2010)104;%%
  %83 citations counted in INSPIRE as of 07 Nov 2018
	
	%\cite{Lunin:2005jy}
\bibitem{Lunin:2005jy}
  O.~Lunin and J.~M.~Maldacena,
 ``Deforming field theories with \text{U}(1) x \text{U}(1) global symmetry and their gravity duals,''
  JHEP {\bf 0505} (2005) 033
  doi:10.1088/1126-6708/2005/05/033
  [hep-th/0502086].
  %%CITATION = doi:10.1088/1126-6708/2005/05/033;%%
  %518 citations counted in INSPIRE as of 07 Nov 2018
	
	%\cite{OColgain:2010ev}
\bibitem{OColgain:2010ev}
  E.~O Colgain, J.~B.~Wu and H.~Yavartanoo,
  ``On the generality of the LLM geometries in M-theory,''
  JHEP {\bf 1104} (2011) 002
  doi:10.1007/JHEP04(2011)002
  [arXiv:1010.5982 [hep-th]].
  %%CITATION = doi:10.1007/JHEP04(2011)002;%%
  %18 citations counted in INSPIRE as of 26 Nov 2018
	

%\cite{Youm:1999zs}
\bibitem{Youm:1999zs}
  D.~Youm,
 ``Partially localized intersecting BPS branes,''
  Nucl.\ Phys.\ B {\bf 556} (1999) 222
  doi:10.1016/S0550-3213(99)00384-3
  [hep-th/9902208].
  %%CITATION = doi:10.1016/S0550-3213(99)00384-3;%%
  %76 citations counted in INSPIRE as of 23 Nov 2018
	
	
	%\cite{Imamura:2001cr}
\bibitem{Imamura:2001cr}
  Y.~Imamura,
  ``1/4 BPS solutions in massive IIA supergravity,''
  Prog.\ Theor.\ Phys.\  {\bf 106} (2001) 653
  doi:10.1143/PTP.106.653
  [hep-th/0105263].
  %%CITATION = doi:10.1143/PTP.106.653;%%
  %25 citations counted in INSPIRE as of 23 Nov 2018
  
\bibitem{Jarv:2000zv}
L.~Jarv and C.~V.~Johnson,
``Orientifolds, M theory, and the ABCD's of the enhancon,''
Phys.\ Rev.\ D {\bf 62} (2000) 126010
doi:10.1103/PhysRevD.62.126010
[hep-th/0002244].
%%CITATION = doi:10.1103/PhysRevD.62.126010;%%
%20 citations counted in INSPIRE as of 27 Nov 2018

	
\end{thebibliography}
\end{document}